\documentclass{article}
\usepackage[utf8]{inputenc}
\usepackage[english]{babel}
\usepackage{authblk}
\usepackage{amsmath}
\usepackage{setspace}
\usepackage[top = 2.5cm, bottom = 2.5cm, outer = 2.5cm, inner = 2.5cm]{geometry}
\usepackage{hyperref}
\usepackage{multicol}
\usepackage{cite}
\usepackage{textcomp}
\usepackage{graphicx}
\usepackage[font=small]{caption, subcaption} 
\usepackage{color}
\usepackage{textcomp}
\usepackage{doi}
\usepackage{booktabs}

\date{}
\begin{document}

\title{Intrinsic RESET speed limit of valence change memories}
\author[1,2,*]{M. von Witzleben}
\author[2]{S. Wiefels}
\author[1]{A. Kindsmüller}
\author[1]{P. Stasner}
\author[1]{F. Berg}
\author[3]{F. Cüppers}
\author[3]{S.~Hoffmann-Eifert}
\author[1,2,3]{R. Waser}
\author[2]{S. Menzel}
\author[1]{U. Böttger}
\affil[1]{Institut für Werkstoffe der Elektrotechnik 2, RWTH Aachen University, Sommerfeldstraße 24, D-52074 Aachen, Germany}
\affil[2]{Peter Grünberg Institut 7, Forschungszentrum Jülich and JARA-FIT, Wilhelm Johnen Straße, D-52428 Jülich, Germany}
\affil[3]{Peter Grünberg Institut 10, Forschungszentrum Jülich and JARA-FIT, Wilhelm Johnen Straße, D-52428 Jülich, Germany}
\affil[*]{Corresponding author: \href{mailto:witzleben@iwe.rwth-aachen.de}{witzleben@iwe.rwth-aachen.de}}

\maketitle
\begin{abstract}
During the last decade, valence change memory (VCM) has been extensively studied due to its promising features, such as a high endurance and fast switching times. The information is stored in a high resistive state (logcial~`0', HRS) and a low resistive state (logcial~`1', LRS). It can also be operated in two different writing schemes, namely a unipolar switching mode (LRS and HRS are written at the same voltage polarity) and a bipolar switching mode (LRS and HRS are written at opposite voltage polarities). VCM, however, still suffers from a large variability during writing operations and also faults occur, which are not yet fully understood and, therefore, require a better understanding of the underlying fault mechanisms. In this study, a new intrinsic failure mechanism is identified, which prohibits RESET times (transition from LRS to HRS) faster than 400\,ps and possibly also limits the endurance. We demonstrate this RESET speed limitation by measuring the RESET kinetics of two valence change memory devices (namely Pt/TaO$_\mathrm{x}$/Ta and Pt/ZrO$_\mathrm{x}$/Ta) in the time regime from 50\,ns to 50\,ps, corresponding to the fastest writing time reported for VCM. Faster RESET times were achieved by increasing the applied pulse voltage. Above a voltage threshold it was, however, no longer possible to reset both types of  devices. Instead a unipolar SET (transition from HRS to LRS) event occurred, preventing faster RESET times.  The occurrence of the unipolar SET is attributed to an oxygen exchange at the interface to the Pt~electrode, which can be suppressed by introducing an oxygen blocking layer at this interface, which also allowed for 50\,ps fast RESET times.
\end{abstract}

\section{Introduction}
Among other emerging memory technologies, valence change memory (VCM) can not only be used as binary storage class memory, but also to perform in-memory calculations or to realize neuromorphic applications~\cite{Zidan2018001, Dittmann2019001, Ielmini2020001, Zhang2020002}. In the binary mode, the information is encoded in a high and a low resistive state (HRS and LRS). These two states can be programmed with electrical stimuli. The transition from the HRS to LRS is referred to as SET and the opposite transition as RESET. A VCM device usually consists of a mixed electronic-ionic conducting layer (e.g.~TaO$_\mathrm{x}$ or~ZrO$_\mathrm{x}$), sandwiched between two asymmetric metallic electrodes \cite{Waser2009004, Yang2013001, Ielmini2016002, Chen2020001}. One of the two electrodes is inert (e.g.~Pt) and referred to as active electrode, whereas the opposite electrode is oxygen affine (e.g.~Ta). The devices are programmed by applying electrical stimuli, during which an n-conducting filament consisting of mobile donors is either formed (SET) or ruptured (RESET).  Spectroscopic studies have identified these mobile donors as oxygen vacancies \cite{Baeumer2015003, Skaja2015001, Yalon2015001, Kindsmueller2018001, Martino2020001}. Especially for TaO$_\mathrm{x}$-based devices also a movement of metallic cations was observed \cite{Wedig2015001, Ma2019002, Rosario2019001}.  During the SET, a negative voltage is applied to the active electrode (attracting mobile donors) and a positive voltage during the RESET (repelling mobile donors). As the voltages required for the SET and RESET have opposite signs, this corresponds to a bipolar switching mode.

For the binary switching mode, a high endurance of up to 10$^{12}$~cycles \cite{Lee2011023, Li2017016} and writing times below 1\,ns\cite{Choi2013002, Choi2016002, Torrezan2011001, Wang2017006, Pickett2012001, Lee2010015} have been reported. A successful market launch also requires high storage densities and, consequently, the integration of VCM devices into 3D structures \cite{Lin2020001}. Due to the presence of Joule heating during the SET \cite{Yalon2012001, Yalon2015002} and during the RESET operation \cite{Marchewka2015002}, the writing time depends strongly non-linearly on the applied voltage\cite{Menzel2011001, Witzleben2017001}. This non-linearity allows VCM devices to overcome the voltage-time-dilemma\cite{Menzel2015001}, which describes the necessity for high data retention during the read-out (at low voltages) and fast writing times (at high voltages) \cite{Strukov2009001}. Programming VCM devices with shorter electrical stimuli also enhances the devices' endurance \cite{Chen2012008, Chen2015008}. Calculations and simulations show that for nanoscale devices faster writing times of 1\,ps are possible \cite{Menzel2018004}. Sub-nanosecond switching times are still realizable, if the VCM devices were integrated into nanoscale crossbar array structures \cite{Fouda2018001}. Also, first neuromorphic applications have been demonstrated on a sub-nanosecond timescale: Ma~et~al. recently demonstrated spike timing-dependent plasticity (STDP) in memristive devices with 600\,ps pulses \cite{Ma2020001}. Concepts of GHz deep neural networks were developed, which are potentially faster than today's CPUs and GPUs by a factor of 30,000 \cite{Gokmen2016001}. Most applications in the GHz regime require, consequently, fast writing times. So far, the speed limit was always attributed to the electrical charging of the devices, assuming that the intrinsic speed limit is only limited by the attempt frequency of the mobile donors in the THz~domain \cite{Menzel2018004}.

Recently, we have shown that 50\,ps fast SET times can be achieved for TaO$_\mathrm{x}$- and ZrO$_\mathrm{x}$-based VCM~devices \cite{Witzleben2020001}, which corresponds to the limitation of the experimental setup. We also demonstrated that the SET kinetics are mainly delayed by the electrical charging time in the sub-nanosecond regime and not by intrinsic processes, such as the migration of ions or the heating of the filamentary region \cite{Witzleben2021001}. The fast heating results from the narrow filament, which has only a diameter in the range from 1\,nm to 3\,nm \cite{Privitera2015001}. While the SET kinetics have been studied quite extensively, studies investigating possible intrinsic RESET speed limits are rare. Most studies show only 10 or fewer successful RESET operations \cite{Torrezan2011001, Choi2013002, Choi2016002} and are not discussing possible RESET speed limitations. Only Wang et al. studied RESET kinetics in dependence on the applied voltage in the sub-nanosecond regime and observed a lower change in resistance at shorter pulse widths than 800\,ps \cite{Wang2017006}. At pulse widths below 200\,ps the VCM device switched randomly between the HRS and LRS. They explain this observation with a lower heating of the filamentary region during the RESET at shorter pulse widths than 800\,ps. Their argumentation, however, does not account for the random switching between the LRS and the HRS.

In this study, we propose a different failure mechanism limiting the RESET kinetics in the sub-nanosecond regime: For this purpose, we focus on the RESET kinetics in the regime from 50\,ns to 50\,ps and show that they are intrinsically limited by the presence of a unipolar switching mode \cite{Ielmini2011003, Prakash2013001}. Different to previous studies, we also acquired a much larger dataset to demonstrate the reproducibility of successful RESET operations. At slower timescales (above 700\,ps) the RESET kinetics depend exponentially on the applied voltage, which was already demonstrated in other studies \cite{Marchewka2015002, Lu2015001, Fleck2016003, Schoenhals2017004, Cueppers2019004}. By increasing the applied voltage we could successfully reset the TaO$_\mathrm{x}$-based device within 670\,ps at a voltage of 1.6\,V and the ZrO$_\mathrm{x}$-based device within 480\,ps at an voltage of 1.8\,V. At higher voltages, the devices could not be driven to the HRS. Instead of an increase in resistance, a unipolar SET could be observed, decreasing the device's resistance and, thereby, preventing faster RESET times. 

This unipolar switching mode has already been observed for TaO$_\mathrm{x}$- \cite{Gao2018002} and ZrO$_\mathrm{x}$-based devices \cite{Lin2016002}, but has never been considered as failure meachanism for neither the RESET kinetics, nor the endurance. It can be triggered by applying a positive voltage to the active electrode, with a higher amplitude compared to the RESET. This could result in a higher heating of the filamentary region than during the bipolar SET and RESET, which in turn could initiate thermo-diffusion of the mobile donors (also referred to as thermophoresis) \cite{Strukov2012001}. To protect the device from damage, a current compliance is crucial during the unipolar SET, which also exhibits abrupt threshold switching. This switching mode was strongly investigated in the years up to 2013 in the hope of achieving a similar performance as for the bipolar switching mode. However, the current compliance to achieve the unipolar SET, has to be in the range of 1\,mA \cite{Yanagida2013001}, which increases the power consumption and the devices' degradation during cycling. To our knowledge, the highest measured endurance of a unipolar switching mode amounts to only 10$^6$~cycles \cite{Baek2004003}, being far worse than the endurance of the bipolar switching mode (10$^{12}$~cycles \cite{Lee2011023}). As this mode prohibits faster RESET times and only achieves a poor performance with regard to power consumption and endurance, and also limits the writing time, it constitutes a failure mechanism.

The kinetics of this unipolar SET are also investigated in the time regime from 250\,ps to 50\,ps. We demonstrate that this unipolar SET event can be conducted within 50\,ps. From the RESET and unipolar SET kinetics, we derive voltage programming windows, from which the intrinsic RESET speed limitation can be derived. Finally, we suppressed the oxygen exchange at the Pt~electrodes with a 1.0\,nm thin Al$_2$O$_3$~layer, which also allowed for 50\,ps fast RESET times for both devices, which is \-- to our knowledge \-- the fastest reported RESET time of redox based random access memories (ReRAMs).

\section{RESET kinetics}

The RESET and unipolar SET kinetics measurements were conducted on a Pt/TaO$_\mathrm{x}$/Ta and a Pt/ZrO$_\mathrm{x}$/Ta device. Both devices have a size of $2 \times 2$\,\textmu m$^2$ and were also used in \cite{Witzleben2020001}, showing that the SET operation can occur within 50\,ps. The 30\,nm thick Pt~bottom electrode serves as active electrode. The film thickness of the TaO$_\mathrm{x}$ and the ZrO$_\mathrm{x}$ amounts to 5\,nm and the film thickness of the Ta~top electrode to 20\,nm. Further information are given in the methods section. In recent publications, we have shown that devices with these material stacks have an endurance of at least 10$^6$~cycles \cite{Kim2016006, Wiefels2020004}.

\begin{figure}[ht]
\begin{subfigure}{0.49\textwidth}
		\centering
		\includegraphics[scale = 1]{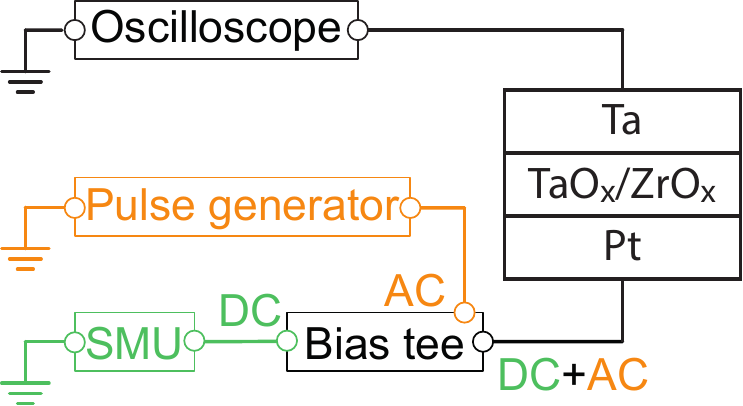}
		\subcaption{}
	\end{subfigure}
	\begin{subfigure}{0.49\textwidth}
		\centering
		\includegraphics[scale = 1]{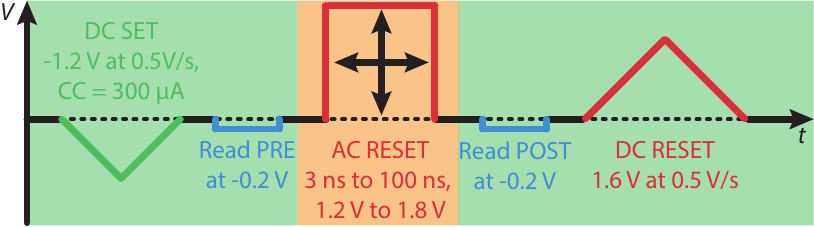}
		\subcaption{}
	\end{subfigure}\\
	\vspace{4mm}\\
	\begin{subfigure}{0.49\textwidth}
		\centering
		\includegraphics[scale = 0.9]{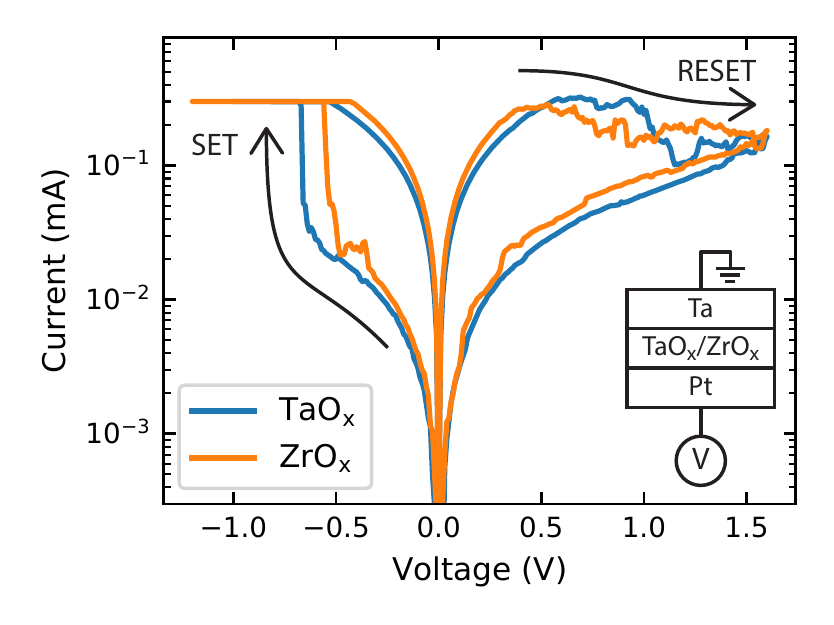}
		\subcaption{}
	\end{subfigure}
\caption{(a) Illustration of experimental setup. The pulse generator is connected to the AC port (orange) of the bias tee and the SMU to the DC~port (green). The combined output of the bias tee (DC + AC) is connected to the Pt~bottom electrode of the VCM device. The current response is measured with an oscilloscope, connected at the Ta~top electrode. (b) Measurement procedure to determine the RESET kinetics in the range from 50\,ns to 400\,ps. Read operations are indicated in blue, SET operations in green and RESET operations in red. All voltages are applied to the Pt~active electrode. The green and orange shaded areas mark the part measured at the DC and the AC~port of the bias tee, respectively. (c) \textit{I(V)} characteristics of both devices (sweep rate: 0.5\,V/s).  Reprinted from \cite{Witzleben2020001}, with the permission of AIP Publishing.}
\label{fig1}
\end{figure}

The experimental setup is sketched in Fig.~\ref{fig1}(a) and is explained in detail in \cite{Witzleben2020001}. A pulse generator is connected to the alternating current (AC) port and a source measure unit (SMU) to the direct current (DC) port of a broadband bias tee. The combined DC + AC port of the bias tee is connected to the Pt~bottom electrode of the VCM device, to which all indicated voltages in this paper were applied. Finally, the current response is measured with a real-time oscilloscope at the Ta~top electrode. More information on the experimental setup is given in the methods section. The \textit{I(V)}-characteristics of both devices are shown in Fig.~\ref{fig1}(c).

To achieve proper impedance matching, both devices were integrated into coplanar waveguide (CPW) structures, consisting of three parallel stripes  (ground-signal-ground, GSG). These CPW structures have gained more attention in recent years, as they can be used to realize radio frequency (RF) switches with memristive devices \cite{Pi2015001, Wainstein2021002}. Recently, we have shown for the TaO$_\mathrm{x}$-based device that the electrical charging occurs within less than 80\,ps, if the device is integrated in a proper CPW structure \cite{Witzleben2021001}. The scattering parameters of both devices are shown in the supplementary Fig.~\ref{sfig1}. For these scattering parameters the devices' charging times were derived in the supplementary Fig.~\ref{sfig2}, showing that the ZrO$_\mathrm{x}$-based device can even be charged within less than 70\,ps.

The measurement procedure to determine the RESET kinetics on the timescale from 50\,ns to 400\,ps is depicted in Fig.~\ref{fig1}(b). At the beginning of every cycle, the devices were driven to the LRS (ranging from 1.0\,k$\Omega$ to 3.0\,k$\Omega$) by using a voltage sweep  with an amplitude of -1.2\,V at a sweep rate of 0.5\,V/s. To protect the devices from damage a current compliance of 100\,µA was used. The devices' resistance was measured at a voltage of -0.2\,V before and after the application of the pulse and is referred to as $R_\mathrm{PRE}$ and $R_\mathrm{POST}$, respectively. The RESET pulses' amplitudes were adjusted between 1.2\,V and 1.6\,V\footnote{This is the effective pulse voltage seen by the device. The pulses emitted by the pulse generator only have half the amplitude. As the 50\,$\Omega$ transmission line is terminated with a high ohmic VCM device (even the resistance of the LRS is much larger than 50\,$\Omega$), the voltage at the device effectively doubles. This is due to the overlap of incoming pulse and the reflected pulse at the VCM~device (see also supplementary section~\ref{sec:charging_time} or \cite{Witzleben2021001})}. At the end of each cycle, the device was driven to the HRS by applying a voltage sweep with an amplitude of 1.6\,V at a sweep rate of 0.5\,V/s.  The pulse width was reduced with every increase in amplitude to reduce the stress on the device.

\begin{figure}[ht]
\begin{subfigure}{0.49\textwidth}
		\centering
		\includegraphics[scale = 1]{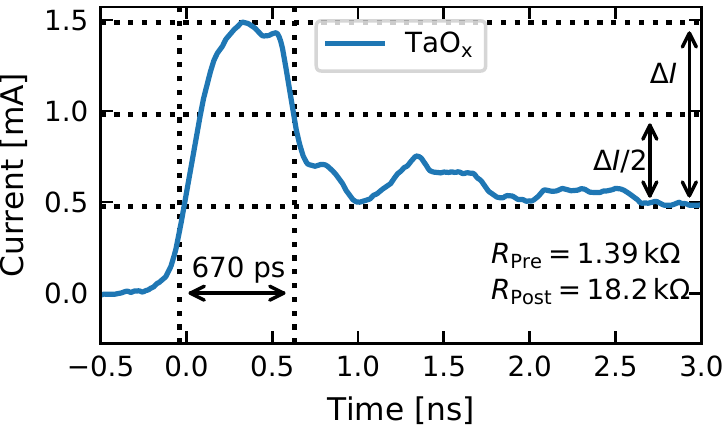}
		\subcaption{}
	\end{subfigure}
	\begin{subfigure}{0.49\textwidth}
		\centering
		\includegraphics[scale = 1]{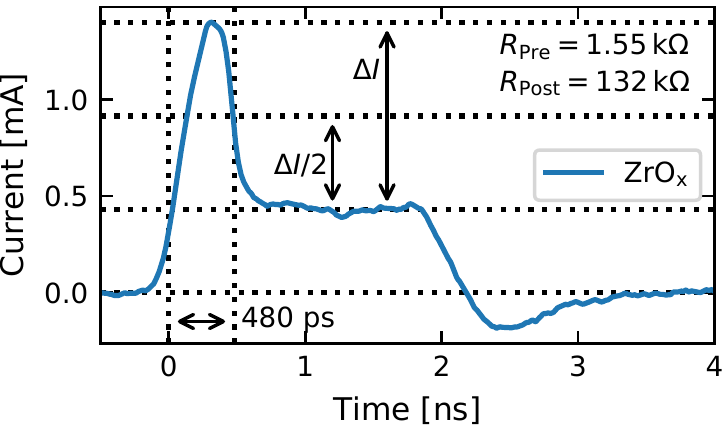}
		\subcaption{}
	\end{subfigure}\\
	\vspace{4mm}\\
	\begin{subfigure}{0.49\textwidth}
		\centering
		\includegraphics[scale = 1]{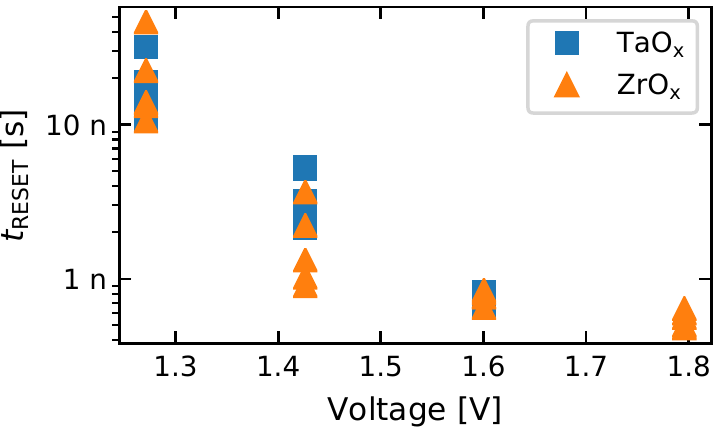}
		\subcaption{}
	\end{subfigure}
\caption{(a) Current response to a pulse with an amplitude of 1.6\,V, applied to the TaO$_\mathrm{x}$-based device. The horizontal dotted lines mark the maximum current, the half value current, and the minimum current (top to bottom). The determined RESET time is denoted by the vertical dotted lines. On the lower right, the change in resistance is indicated. (b) Current response to a pulse with an amplitude of 1.8\,V, applied to the ZrO$_\mathrm{x}$-based device. The additional horizontal dotted line indicates the zero current baseline. (c) Reset times $t_\mathrm{RESET}$, plotted against the applied voltage.}
\label{fig2}
\end{figure}

Two exemplary current responses to the applied voltage pulse are shown Fig.~\ref{fig2}(a) and~(b) for the TaO$_\mathrm{x}$ and ZrO$_\mathrm{x}$-based device, respectively. In both cases, the current increases rapidly at the beginning of the pulse and decreases, subsequently, to values between 400\,µA to 500\,µA, which corresponds to the RESET. The contribution of the capacitive current is negligible. This can be seen at the end of the pulse applied to the ZrO$_\mathrm{x}$-based device in Fig.~\ref{fig2}(b). The negative current peak corresponds to the capacitive current and is much smaller than the current peak at the beginning of the pulse. The RESET time $t_\mathrm{RESET}$ is defined as the time difference between the time at which the current surpasses 20\,\% of the maximum current at the beginning of the pulse (first vertical dotted line in Fig.~\ref{fig2}(a) and~(b)) and the time at which the current reaches its half value (second vertical dotted line). The half value of the current $\Delta I$/2 is defined as half the current difference between the minimum and the maximum current $\Delta I$ (see illustrations in Fig.~\ref{fig2}(a) and~(b)). A similar definition was used in our previous publications \cite{Marchewka2015002, Fleck2016003}.

In case of the current response of the TaO$_\mathrm{x}$-based device in Fig.~\ref{fig2}(a), a 10\,ns pulse with an amplitude of 1.6\,V was applied. During this pulse the device's resistance increased from 1.39\,k$\Omega$ to  18.2\,k$\Omega$ and a RESET time of 670\,ps was determined. By using higher pulse amplitudes, faster RESET times could be possible. Increasing the pulse amplitude to 1.8\,V, however, resulted in dielectric breakdown and the device's resistance started to decrease. This is shown in the supplementary Fig.~\ref{sfig3}. All determined RESET times $t_\mathrm{RESET}$ are shown in Fig.~\ref{fig2}(c) in dependence of the applied voltage. Up to a voltage of 1.6\,V, the RESET time depends strongly non-linearly on the applied voltage, which was expected. At a voltage of 1.8\,V (only values for ZrO$_\mathrm{x}$) the kinetics start to bend towards slower RESET times, which is attributed to the influence of the pulse generator's rise time ($\approx$\,360\,ps). 

\begin{figure}[ht]
\begin{subfigure}{0.49\textwidth}
		\centering
		\includegraphics[scale = 1]{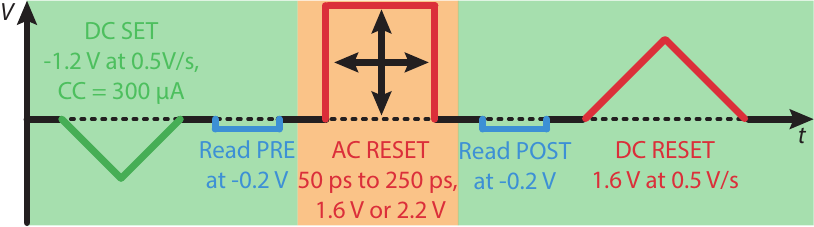}
		\subcaption{}
	\end{subfigure}
	\begin{subfigure}{0.49\textwidth}
		\centering
		\includegraphics[scale = 1]{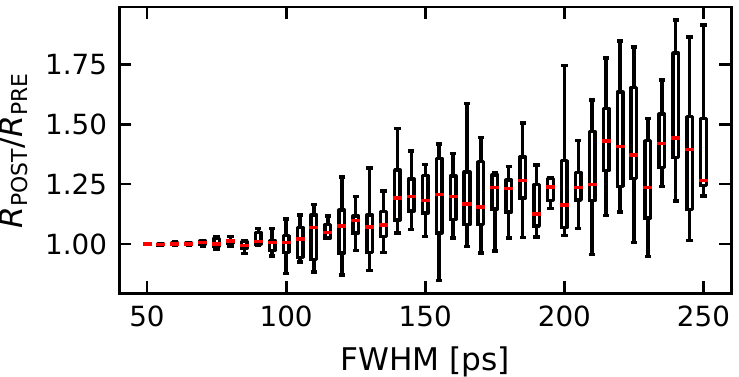}
		\subcaption{}
	\end{subfigure}\\
	\vspace{4mm}\\
	\begin{subfigure}{0.49\textwidth}
		\centering
		\includegraphics[scale = 1]{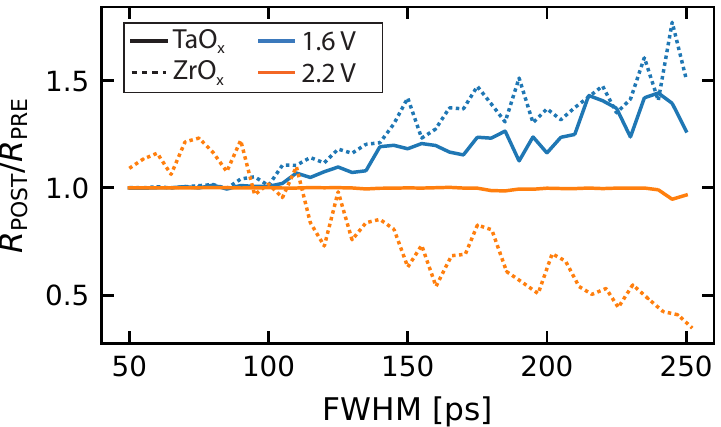}
		\subcaption{}
	\end{subfigure}
\caption{(a) Measurement procedure to determine the RESET kinetics in the range from 250\,ps to 50\,ps. (b) Exemplary result for the TaO$_\mathrm{x}$-based device at a voltage of 1.6\,V. The boxplots indicate the scattering of the ratio $R_\mathrm{POST}$/$R_\mathrm{PRE}$ at a specific full width half maximum (FWHM). The red bar indicates the median value. (c) Median values of all measurements. The solid and dotted lines indicate the results of the TaO$_\mathrm{x}$ and ZrO$_\mathrm{x}$-based devices, respectively. The pulse amplitude amount to 1.6\,V (blue) and 2.2\,V (orange). The solid blue line in (c) represents the median values of (b).}
\label{fig3}
\end{figure}

To decrease the influence of the pulse generator's rise time, the RESET kinetic measurements in the range from 250\,ps to 50\,ps were conducted with a faster pulse generator, having a rise time of only $\approx$\,35\,ps. The measurement cycle is sketched in Fig.~\ref{fig3}(a), and is almost identical to the previous (Fig.~\ref{fig1}(b)), only the applied RESET pulse has changed. The pulse amplitude was once chosen to 1.6\,V and once to 2.2\,V.  As in~\cite{Witzleben2020001}, the pulse width was increased from 50\,ps to 250\,ps in steps of 5\,ps. Each cycle was repeated 10~times\footnote{The measurement cycles of the ZrO$_\mathrm{x}$-based device at a voltage of 1.6\,V were repeated 20~times to achieve a smooth line of $R_\mathrm{Post}$/$R_\mathrm{PRE}$.}. 


An exemplary measurement on the TaO$_\mathrm{x}$-based device at a voltage of 1.6\,V is shown in Fig.~\ref{fig3}(b). The ratio of the resistance before $R_\mathrm{PRE}$ and after $R_\mathrm{POST}$ the pulse's application is plotted as boxplot against the FWHM of the applied pulse. The red bar marks the median. As expected, the ratio $R_\mathrm{POST}$/$R_\mathrm{PRE}$ increases with increasing FWHM. The median of $R_\mathrm{POST}$/$R_\mathrm{PRE}$, however, remains below~1.5 for all FWHMs, which means that no successful RESET occurs in this time interval\footnote{For a successful RESET operation the ratio  of $R_\mathrm{POST}$/$R_\mathrm{PRE}$ should be greater than 3 \cite{Wiefels2020004}}. The median values of $R_\mathrm{POST}$/$R_\mathrm{PRE}$ of all four measurements are shown in Fig.~\ref{fig3}(c). The corresponding boxplots can be found in the supplementary Fig.~\ref{fig:RESET_ratio}. As the ratio $R_\mathrm{POST}$/$R_\mathrm{PRE}$ does not yield any information about the absolute resistance values, the $R_\mathrm{POST}$ values are shown in the supplementary Fig.~\ref{fig:RESET_POST}. The initial resistances $R_\mathrm{PRE}$ were always in the range from 1.0\,k$\Omega$ to 2.5\,k$\Omega$.

At 1.6\,V, the ratio $R_\mathrm{POST}$/$R_\mathrm{PRE}$ of the ZrO$_\mathrm{x}$-based device (dashed blue line in Fig.~\ref{fig3}(c)) also remains below 2.0 for all FWHMs, indicating that the ZrO$_\mathrm{x}$-based device cannot be driven to HRS within 250\,ps. At an amplitude of 2.2\,V, $R_\mathrm{POST}$/$R_\mathrm{PRE}$ of the TaO$_\mathrm{x}$-based device (solid orange line in Fig.~\ref{fig3}(c)) remained near unity for all FWHMs, while $R_\mathrm{POST}$/$R_\mathrm{PRE}$ of the ZrO$_\mathrm{x}$-based device (dotted orange line in Fig.~\ref{fig3}(c)) decreased with increasing FWHM. Consequently, it is not possible with both devices to achieve RESET times faster than 250\,ps by increasing the pulse amplitude. The decrease in resistance of the ZrO$_\mathrm{x}$-based device at 2.2\,V indicates the presence of a unipolar switching mode. 

\section{Unipolar SET kinetics}

\begin{figure}[ht]
	\begin{subfigure}{0.49\textwidth}
		\centering
		\includegraphics[scale = 1]{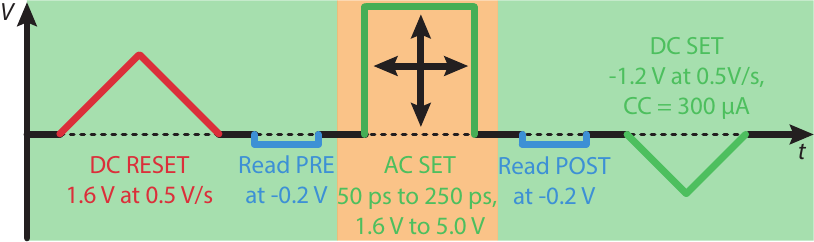}
		\subcaption{}
	\end{subfigure}
	\begin{subfigure}{0.49\textwidth}
		\centering
		\includegraphics[scale = 1]{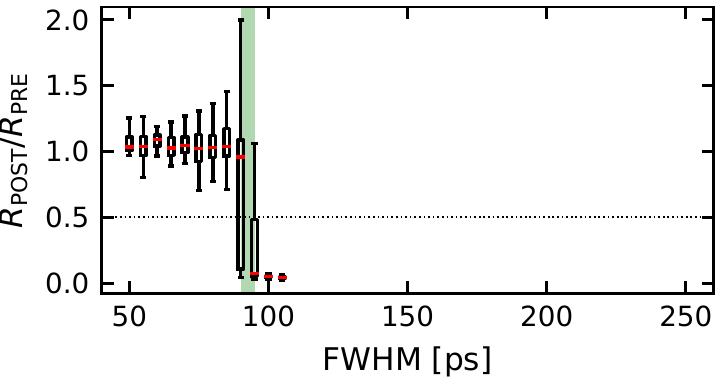}
		\subcaption{}
	\end{subfigure}\\
	\vspace{4mm}\\
	\begin{subfigure}{0.49\textwidth}
		\centering
		\includegraphics[scale = 1]{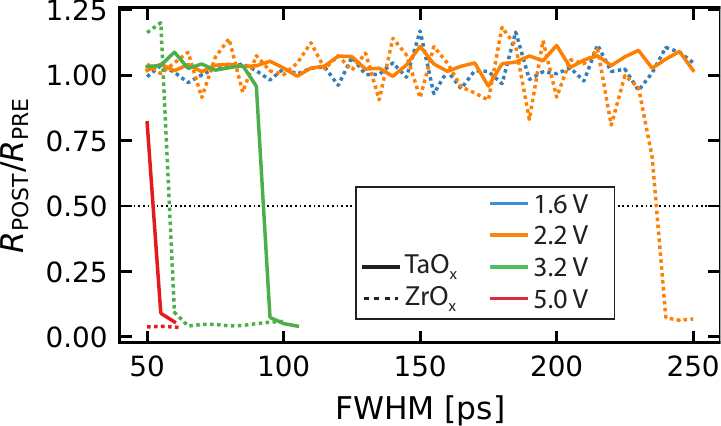}
		\subcaption{}
	\end{subfigure}
	\begin{subfigure}{0.49\textwidth}
		\centering
		\includegraphics[scale = 1]{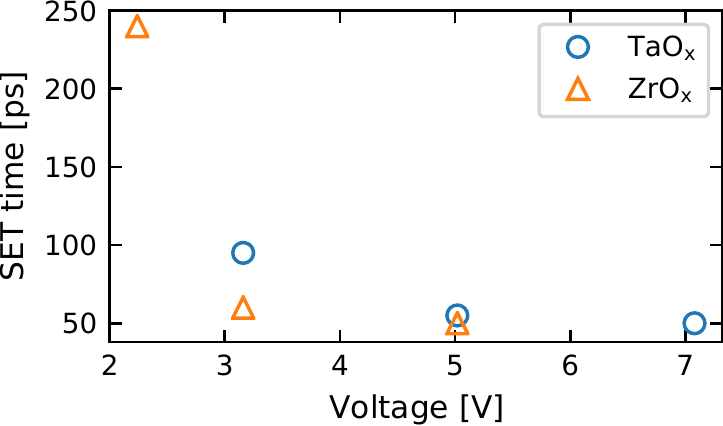}
		\subcaption{}
	\end{subfigure}
\caption{(a) Measurement procedure to determine the unipolar SET kinetics in the range from 250\,ps to 50\,ps. (b) Exemplary result for the TaO$_\mathrm{x}$-based device at a voltage of 3.2\,V.  (c) Median values of all measurements. The solid and dotted lines indicate the results of the TaO$_\mathrm{x}$ and ZrO$_\mathrm{x}$-based devices, respectively. The pulse amplitude amount to 1.6\,V (blue), 2.2\,V (orange), 3.2\,V (green), and 5.0\,V (red). The solid green line in (c) represents the median values of (b).}
\label{fig4}
\end{figure}

The measurement procedure to determine the unipolar SET kinetics is depicted in Fig.~\ref{fig4}(a). Different to the measurement procedures of the RESET kinetics, the device is driven to the HRS at the beginning of the measurement cycle by applying a voltage sweep with an amplitude of 1.6\,V. The resulting resistance values were again in the range from 10\,k$\Omega$ to 30\,k$\Omega$. At the end, it is driven to the LRS with a voltage sweep of -1.2\,V. The sweep rate (0.5\,V/s) and the current compliance during the SET sweep (300\,\textmu A) are identical to the values chosen in the RESET kinetics. The pulse amplitude was again varied from 50\,ps to 250\,ps and the amplitude from 1.6\,V to 5.0\,V. As the unipolar switching mode has a large variabilty for TaO$_\mathrm{x}$- \cite{Sakotsubo2010001} and ZrO$_\mathrm{x}$-based devices \cite{Wu2007004, Lin2016002}, this cycle was repeated at least 10~times\footnote{The exact numbers are listed in Table~\ref{tab:no_cycles} of the supplementary information.}, until smooth curves were achieved for $R_\mathrm{POST}$/$R_\mathrm{PRE}$.

An exemplary measurements is shown in Fig.~\ref{fig4}(b), in which voltage pulses with an amplitude of 3.2\,V were applied to the TaO$_\mathrm{x}$-based device. A short FWHMs, the ratio $R_\mathrm{POST}$/$R_\mathrm{PRE}$ remains near unity, indicating that the resistance of the device does not change during the pulse's application. At FWHMs of about 95\,ps, $R_\mathrm{POST}$/$R_\mathrm{PRE}$ drops towards zero, indicating that the device was successfully driven to the LRS by applying a positive voltage pulse to the active Pt~electrode. As the device is also driven to the HRS with a positive voltage pulse, this corresponds to a unipolar SET. At longer FWHMs, the measurement was aborted to prevent the device from damage. 

This measurement procedure was repeated at different amplitudes (from 1.6\,V to 5.0\,V) for the TaO$_\mathrm{x}$- and ZrO$_\mathrm{x}$-based devices. The median values are plotted in Fig.~\ref{fig4}(c). The boxplot representation of $R_\mathrm{POST}$/$R_\mathrm{PRE}$ and  the absolute values of $R_\mathrm{POST}$ are shown in the supplement Fig.~\ref{fig:SET_ratio_Ta}-\ref{fig:SET_ratio_Zr_POST}. Similar to our previous study, we defined the unipolar SET time, as the time, at which $R_\mathrm{POST}$/$R_\mathrm{PRE}$ reaches a value below 0.5 \cite{Witzleben2020001}. This threshold is indicated as horizontal dashed line in Fig.~\ref{fig4}(c). 

The resulting unipolar SET times are shown in Fig.~\ref{fig4}(d) in dependence of the pulse voltage. The unipolar SET kinetics have similar fast switching times compared to the bipolar SET kinetics from \cite{Witzleben2020001}. Only the voltage has been shifted to higher absolute values. The fastest unipolar SET~time amounts to 50\,ps, which is \-- to our knowledge \--  the fastest unipolar SET measured in ReRAM devices. The fastest reported unipolar SET time, so far, amounts to 16\,ns \cite{Cagli2009001}. As the TaO$_\mathrm{x}$-based device did not switch at a pulse width of 50\,ps an amplitude of 5.0\,V to the LRS, a single measurement cycle with a pulse width of 50\,ps and an amplitude of 7.1\,V was repeated, during which the TaO$_\mathrm{x}$-based device switched to the LRS. Exemplary measured current transients, during which a unipolar SET time of 50\,ps was realized, are shown in Fig.~\ref{fig:SET_50ps} of the supplementary information for both devices. The endurance of the unipolar switching mode was also tested. The TaO$_\mathrm{x}$ and the ZrO$_\mathrm{x}$-based device reached an endurance of 10$^4$~cycles and 3590~cycles, respectively. The description of the measurement procedure and the results are shown in the supplementary~Fig.~\ref{fig:SET_endurance}. Usually, current compliances are required to realize unipolar switching modes \cite{Yanagida2013001}. In this study, the FWHM of the electrical stimuli is much shorter than state-of-the-art current compliances (often operating at time-scales above 1\,\textmu s\cite{Hennen2021001}) and, therefore, the unipolar switching mode can be realized without current compliance.  

\section{Intrinsic RESET speed limitation}

Although fast switching times are usually considered as promising feature, our interpretation of the fast unipolar SET time is that it should be considered as failure mechanism. At higher voltage amplitudes, the unipolar SET occurs faster than the RESET and, thereby, prohibits fast bipolar switching in both directions. This RESET speed limitation is illustrated in the following with RESET programming windows. These windows were estimated with the results from the RESET and the unipolar SET kinetics. As mentioned in the introduction, the unipolar switching mode observed in ReRAM devices, has less promising attributes with regards to endurance and switching power compared to the bipolar switching mode.  

\begin{figure}[ht]
\begin{subfigure}{0.49\textwidth}
		\centering
		\includegraphics[scale = 1]{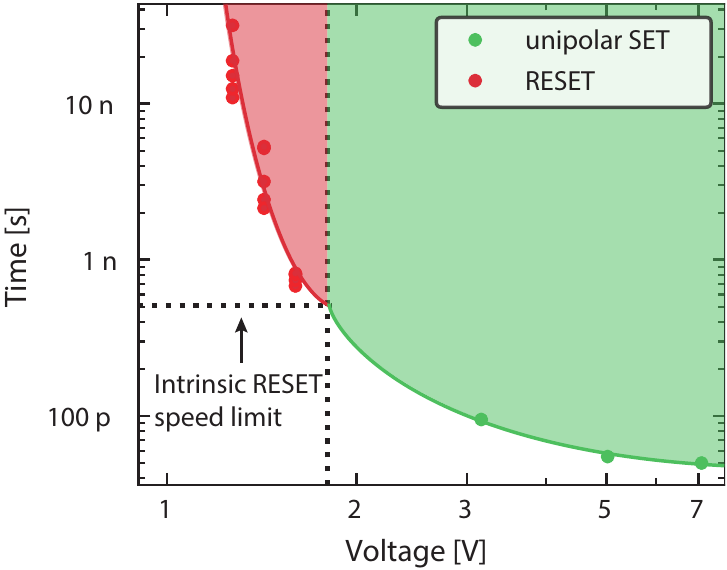}
		\subcaption{TaO$_\mathrm{x}$-based device}
	\end{subfigure}
	\begin{subfigure}{0.49\textwidth}
		\centering
		\includegraphics[scale = 1]{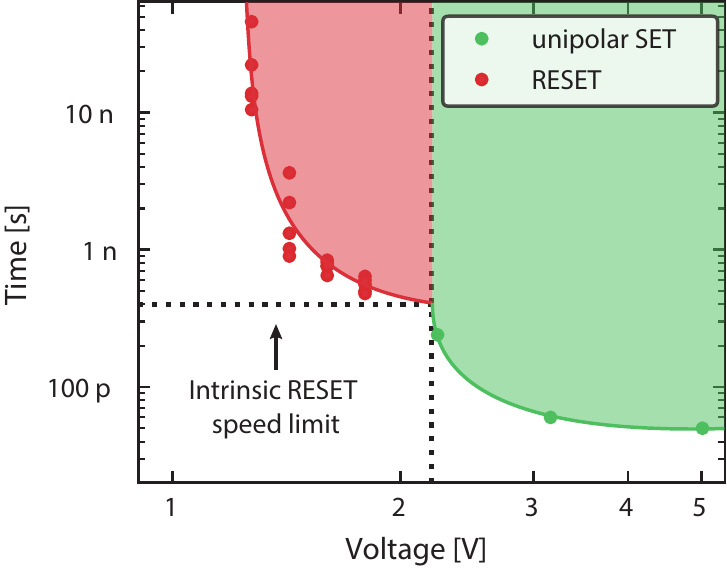}
		\subcaption{ZrO$_\mathrm{x}$-based device}
	\end{subfigure}
	\caption{RESET programming window (a) for the TaO$_\mathrm{x}$ and (b) for the ZrO$_\mathrm{x}$-based device. The red points mark the measured RESET times from Fig.~\ref{fig2}(c) and the green points the measured unipolar SET times from Fig.~\ref{fig4}(d). The red and green areas (drawn by the eye) mark time-voltage combinations, at which either a RESET or a unipolar SET is triggered.}
	\label{fig5}
\end{figure}

The results from the RESET kinetics (Fig.~\ref{fig2}(c)) and the results from the unipolar SET kinetics (Fig.~\ref{fig4}(d)) are plotted in Fig.~\ref{fig5}. The red points mark the measured RESET times and the green points the measured unipolar SET times. The red and green shaded areas mark the voltage-time combinations at which either a RESET or a unipolar SET event is triggered. The green and red lines, encircling the red and green shaded area, are drawn by the eye. Their intersection is an estimate of the fastest possible RESET time, demonstrating that the presence of the unipolar switching mode intrinsically limits faster RESET times. In case of the TaO$_\mathrm{x}$-based device, this intersection occurs at 510\,ps and in case of the ZrO$_\mathrm{x}$-based device at 400\,ps, which marks their intrinsic RESET speed limit. It has, however, to be noted that this RESET programming window will vary from device to device and possibly also from cycle to cycle due to the high variability of the unipolar SET. This might also be the origin of the random switching between the HRS and the LRS in the results of Wang et al. \cite{Wang2017006}. In the sub-100\,ps regime the measured unipolar SET events are also influenced by the electrical charging of the devices. At slower timescales, the unipolar SET is assumed to be time-independent.

\section{Discussion}

From Fig.~\ref{fig5}, two possibilities to achieve faster RESET times can be derived:
\begin{enumerate}
\item Lowering the RESET voltage
\item Suppressing the unipolar switching mode
\end{enumerate}

Torrezan et al. achieved 120\,ps fast RESET times with a TaO$_\mathrm{x}$-based device \cite{Torrezan2011001}. The \textit{I(V)} characteristics of their devices show that the RESET sets in at about 0.35\,V \cite{Strachan2011002}. In contrast, the RESET of the devices used in this study sets in above 0.5\,V (see Fig.~\ref{fig1}(c)). The device stack of both devices is very similar (\cite{Torrezan2011001}: Pt(20\,nm)/TaO$_\mathrm{x}$(7\,nm)/Ta(30\,nm), this work: Pt(30\,nm)/TaO$_\mathrm{x}$(5\,nm)/Ta(20\,nm). We therefore, assume that different fabrication processes are responsible for the lower RESET voltage. 


Suppressing the unipolar SET or mitigating its impact on the bipolar RESET kinetics requires a better understanding of the underlying physical processes. The abrupt nature of the unipolar SET results from a thermal runaway (similar to the bipolar SET) \cite{Ielmini2011003}. This thermal runaway, however, does not necessarily lead to a permanent change in resistance. Especially TaO$_\mathrm{x}$ also exhibits threshold switching \cite{Goodwill2019001}. The sudden current increase, may initiate an oxygen exchange between the oxide and the active Pt~electrode, which leads to a permanent change in the device's resistance. Such an exchange was also observed in the so-called ``eightwise'' switching mode, which has been observed in many transition metal oxides \cite{Cooper2017001, Zhang2018004, Petzold2019001, Siegel2020001}, including TaO$_\mathrm{x}$-based devices \cite{Schoenhals2017004}. This oxygen exchange could also be the origin of the permanent decrease in resistance during the unipolar SET, and could be suppressed in other studies by introducing an oxygen blocking layer such as C \cite{Schoenhals2017004} or  Al$_2$O$_3$ \cite{Zhang2018004, Siegel2020001}.

To test if an oxygen exchange also occurs during the unipolar SET between the oxide and the active Pt~electrode, we introduced a 1.0\,nm thick Al$_2$O$_3$~layer between the active Pt~electrodes and the oxides, resulting in a Pt(30\,nm)/Al$_2$O$_3$(1.0\,nm)/TaO$_\mathrm{x}$(5\,nm)/Ta(20\,nm) and a Pt(30\,nm)/Al$_2$O$_3$(1.0\,nm)/ZrO$_\mathrm{x}$(5\,nm)/Ta(20\,nm) stack (referred to as Al$_2$O$_3$/TaO$_\mathrm{x}$ and Al$_2$O$_3$/ZrO$_\mathrm{x}$ device, respectively). We tested to reset both devices with 50\,ps pulses. The measurement cycle is similar to the one from Fig.~\ref{fig3}(a), only that this time the pulse width was fixed at 50\,ps and the pulse amplitude was chosen to 5.0\,V and 4.0\,V for the Al$_2$O$_3$/TaO$_\mathrm{x}$ and the Al$_2$O$_3$/ZrO$_\mathrm{x}$ device, respectively. This measurement cycle was repeated 100~times for each device.

\begin{figure}[ht]
	\begin{subfigure}{0.49\textwidth}
		\centering
		\includegraphics[scale = 1]{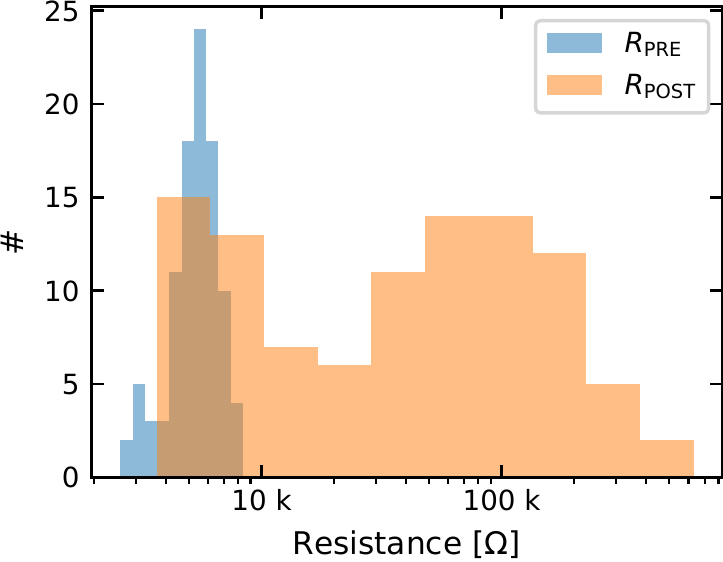}
		\subcaption{Al$_2$O$_3$/TaO$_\mathrm{x}$ device, $V = 5.0$\,V, FWHM = 50\,ps}
	\end{subfigure}
	\begin{subfigure}{0.49\textwidth}
		\centering
		\includegraphics[scale = 1]{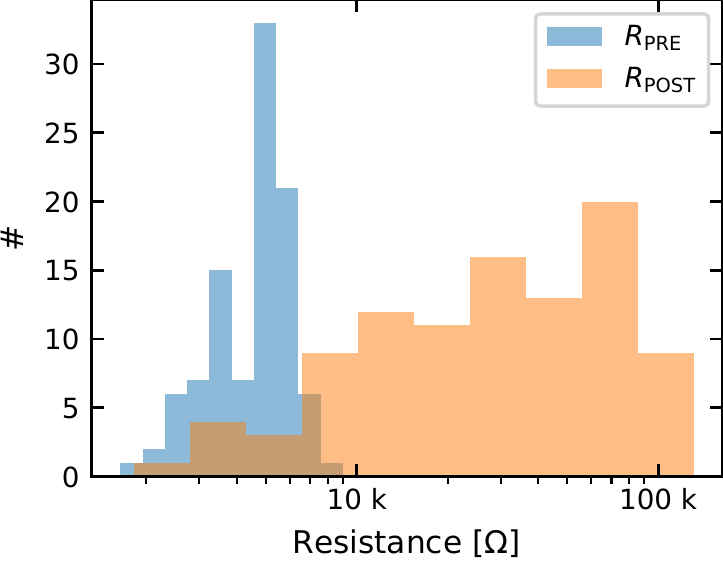}
		\subcaption{Al$_2$O$_3$/ZrO$_\mathrm{x}$ device, $V = 4.0$\,V, FWHM = 50\,ps}
	\end{subfigure}\\
	\vspace{4mm}\\
	\begin{subfigure}{0.49\textwidth}
		\centering
		\includegraphics[scale = 1]{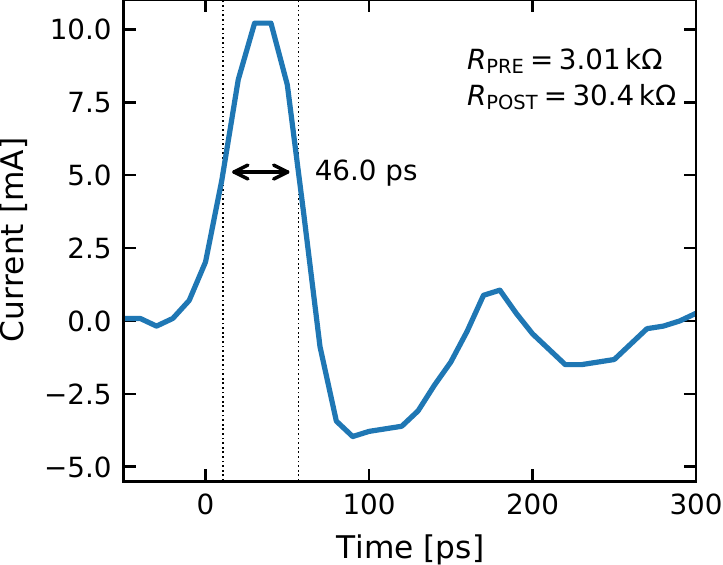}
		\subcaption{Al$_2$O$_3$/TaO$_\mathrm{x}$ device, $V = 5.0$\,V, FWHM = 50\,ps}
	\end{subfigure}
	\begin{subfigure}{0.49\textwidth}
		\centering
		\includegraphics[scale = 1]{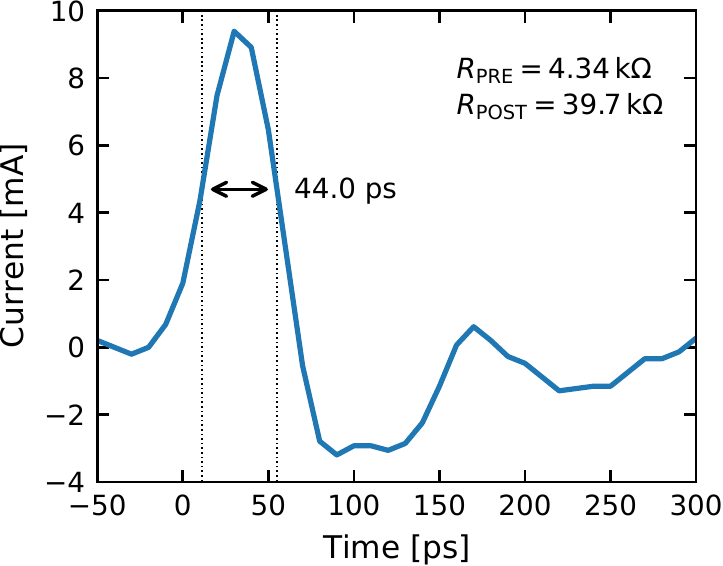}
		\subcaption{Al$_2$O$_3$/ZrO$_\mathrm{x}$ device, $V = 4.0$\,V, FWHM = 50\,ps}
	\end{subfigure}
\caption{Histograms of $R_\mathrm{PRE}$ and $R_\mathrm{POST}$ for the (a) Al$_2$O$_3$/TaO$_\mathrm{x}$ and (b) Al$_2$O$_3$/ZrO$_\mathrm{x}$ device.  Exemplary current responses during the RESET of the (c) Al$_2$O$_3$/TaO$_\mathrm{x}$ and (d) Al$_2$O$_3$/ZrO$_\mathrm{x}$ device. The change in resistance is indicated on the upper right.}
\label{fig6}
\end{figure}

The results for $R_\mathrm{PRE}$ and $R_\mathrm{POST}$ of the Al$_2$O$_3$/TaO$_\mathrm{x}$ and the Al$_2$O$_3$/ZrO$_\mathrm{x}$ device are shown in Fig.~\ref{fig6}(a) and (b), respectively.   During most cycles the devices switched to a higher resistance value after the application, showing that the additional Al$_2$O$_3$~layer improves the feasibility of fast RESET operations. As shown in Fig.~\ref{fig4}(c) the devices without  additional Al$_2$O$_3$ already started to switch to the LRS at a pulse amplitude of 5.0\,V (red lines). To our knowledge, this is the first time that a 50\,ps fast RESET process has been demonstrated for a ReRAM device. Two exemplary current responses are shown in Fig.~\ref{fig6}(c) and (d) for the Al$_2$O$_3$/TaO$_\mathrm{x}$ and Al$_2$O$_3$/ZrO$_\mathrm{x}$ device, respectively. The FHWM of 50\,ps has been preserved in both cases. 

The devices did, however, also remain in the LRS sometimes, indicating that the 1.0\,nm thick Al$_2$O$_3$~layer might only partially prohibit the oxygen exchange at the Pt~active electrode and the remaining oxygen exchange might prohibit successful RESET operations. A similar observation was made by Zhang~et~al., who have placed an Al$_2$O$_3$~layer at the active electrode of a TiO$_\mathrm{x}$-based device to suppress the oxygen exchange at the active interface \cite{Zhang2018004}. Their approach, however, only worked for Al$_2$O$_3$~layers thicker than 2\,nm. Further studies with different oxygen blocking layers or different active electrodes are, consequently, necessary to achieve better control over the RESET process at these fast timescales. 

\begin{figure}[ht]
	\centering
	\includegraphics[scale = 1]{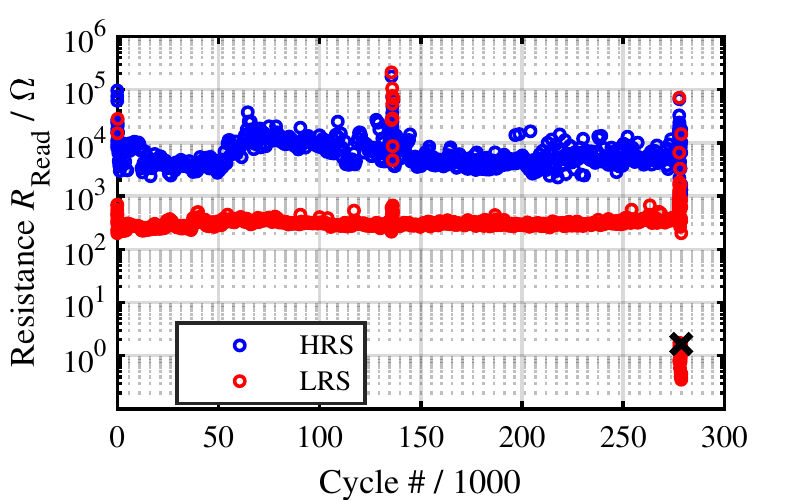}
	\caption{Exemplary endurance measurement on a Pt(30\,nm)/ZrO$_\mathrm{x}$(5\,nm)/Ta(20\,nm) device. The device got irreversibly stuck in the LRS after about 2.8$\cdot 10^{5}$~cycles.}
	\label{fig7}
\end{figure}

Another aspect of the unipolar SET is that it might limit the endurance of VCM based devices. During endurance measurements on ZrO$_\mathrm{x}$ and  TaO$_\mathrm{x}$-based devices, with very similar device fabrication, we observed that the devices were usually trapped in the LRS \cite{Kim2016007, Wiefels2020004}, which might have been triggered by a unipolar SET event. An exemplary endurance measurement from \cite{Wiefels2020004}, during which a ZrO$_\mathrm{x}$-based device got trapped in the LRS is shown in Fig.~\ref{fig7}.   In \cite{Kim2016007} the endurance limitation of TaO$_\mathrm{x}$-based devices was referred to as ``RESET Failure'', which is characterized by an abrupt current increase during a RESET sweep. Afterwards, the devices were stuck in the LRS. This abrupt current increase is also observed during the unipolar SET, only that it is usually controlled with a current compliance \cite{Yanagida2013001}.

\section{Conclusion}

In this study, we have demonstrated that the RESET kinetics of VCM-based devices are intrinsically limited by the coexistance of a unipolar switching mode. Therefore, the RESET kinetics of a TaO$_\mathrm{x}$ and a ZrO$_\mathrm{x}$-based device were investigated in the regime from 50\,ns to 50\,ps. RESET times down to 480\,ps could be achieved by increasing the voltage amplitude of the RESET pulse. At higher amplitudes, a unipolar SET event was triggered which prohibits faster RESET times and rather decreases the devices' resistances. The unipolar SET kinetics were subsequently measured, showing that the unipolar SET can occur within 50\,ps. We attribute the unipolar SET event to a thermal runaway, which followed by an oxygen exchange between the oxide and the active Pt~electrode. This oxygen exchange could be partially suppressed by introducing a 1.0\,nm thick Al$_2$O$_3$~layer between the oxide and the active electrode. With these devices 50\,ps fast RESET operations could be repeated, but only stochastically. We also show, with the data from previous publications, that the occurrence of the unipolar SET limits the endurance of both devices. As the unipolar switching mode has less promising attributes than the bipolar switching mode, the unipolar SET needs to be treated as failure mechanism and further strategies should be developed to suppress its occurrence.

\section*{Methods}

\textbf{Device fabrication and structure}: As substrate served a Si-wafer ($\rho > 10$\,k$\Omega$cm). The top of the wafer was oxidized resulting in a 430\,nm thick SiO$_2$ layer. All deposition steps were conducted by means of RF~magnetron sputtering. The $2 \times 2$\,µm$^2$ devices were structured with optical lithography. More details on the device fabrication can be found in \cite{Schoenhals2015001} and \cite{LaTorre2017001} for the TaO$_\mathrm{x}$- and the ZrO$_\mathrm{x}$-based device, respectively. The 1.0 nm Al$_2$O$_3$ layer was grown by atomic layer deposition (ALD), using trimethylaluminum (TMA) and a remote RF oxygen plasma source \cite{Zhang2018004, Siegel2020001}. This process ensures a uniform and dense Al$_2$O$_3$ layer on the sample surface.\\The devices were formed with a voltage sweep with an amplitude of -4.0\,V at a sweep rate of 0.5\,V/s, applied to the active Pt~electrode. A current compliance of 100\,µA was used during the forming. 
Both devices were integrated into a coplanar waveguide structure, having a length of 590\,µm. The center signal stripe has a width of 100\,µm and a spacing of 60\,µm to the outer grounded stripes. At the center the inner conductor is tapered to 2\,µm, matching the dimensions of the $2 \times 2$\,µm$^2$ devices. More information and illustrations of this structures are given in \cite{Witzleben2021001}. 

\noindent \textbf{Electrical characterization}: The experimental setup is sketched in Fig.~\ref{fig1}. The voltage sweeps and read-outs were conducted with a Keithley~2634B SMU. Two pulse generators were used: pulses with a pulse width above 1.0\,ns were applied with a Picosecond~2600C pulse generator (rise time $\approx 360$\,ps) and pulses between  250\,ps and 50\,ps with a customized pulse generator from the Sympuls~GmbH (rise time $\approx 35$\,ps). It has a fixed pulse amplitude of 5.0\,V (at 50\,$\Omega$), which was adjusted with fixed broadband attenuators. The current responses of the pulse were measured with a Tektronix~DPO73304D real-time oscilloscope, having a bandwidth of 33.0\,GHz and a sample rate of 100\,GS/s. The devices were connected with GSG Z-probes from FormFactor. All components of the setup (except of the oscilloscope) have a bandwidth of 40\,GHz. The measurement procedure is more explicitly described in \cite{Witzleben2020001}.

\section*{Acknowledgements}

This work was supported in part by the Deutsche Forschungsgemeinschaft under Project SFB~917 and in part by the Federal Ministry of Education and Research (BMBF, Germany) in the Project Neuro-inspirierte Technologien der künstlichen Intelligenz (NEUROTEC) under grants 16ES1134 and 16ES1133K. It is based on the Jülich Aachen Research Alliance (JARA-FIT).

\singlespacing

\clearpage

\setcounter{figure}{0}
\setcounter{table}{0}

\makeatletter
\renewcommand \thesection{S\@arabic\c@section}
\renewcommand\thetable{S\@arabic\c@table}
\renewcommand \thefigure{S\@arabic\c@figure}
\makeatother


\appendix
\setcounter{section}{18}
\section{Supplementary Information}
\subsection{Electrical charging time}\label{sec:charging_time}
\begin{figure}[h]
	\begin{subfigure}{0.49\textwidth}
		\centering
		\includegraphics[scale = 1]{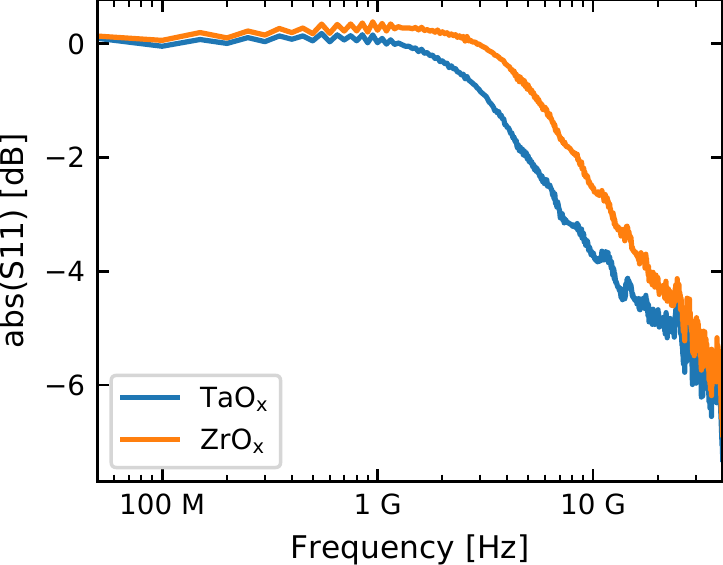}
		\subcaption{}
		\label{fig:dev_s11}
	\end{subfigure}
	\begin{subfigure}{0.49\textwidth}
		\centering
		\includegraphics[scale = 1]{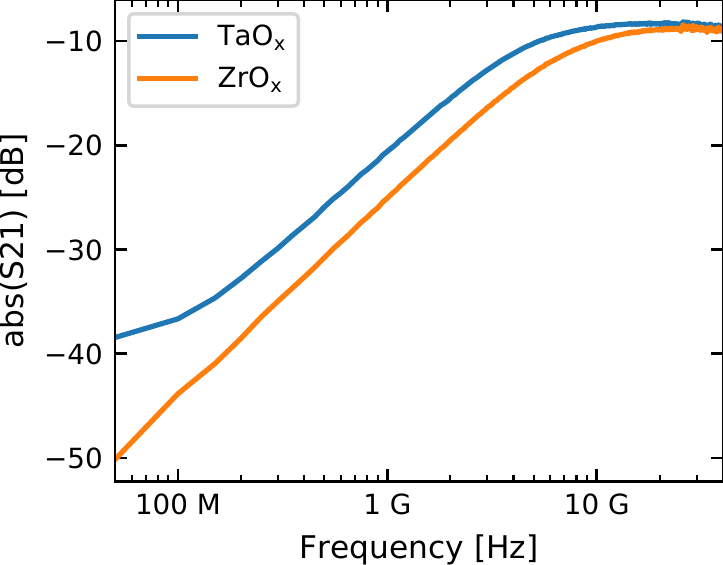}
		\subcaption{}
		\label{fig:dev_s21}
	\end{subfigure}
	\caption{Absolute values of the (a) forward reflection $S_{11}$ and (b) the forward transmission $S_{21}$ of the TaO$_\mathrm{x}$ (blue) and ZrO$_\mathrm{x}$-based (orange) device.}
	\label{sfig1}
\end{figure}

In \cite{Witzleben2021001_S}, we have demonstrated for the TaO$_\mathrm{x}$-based device a charging time of 77.1\,ps by using its scattering parameters. In Fig.~\ref{sfig1}, we have plotted the scattering parameters of both, the TaO$_\mathrm{x}$ and ZrO$_\mathrm{x}$-based device. The forward reflection $S_{11}$ is shown in~(a) and the forward transmission $S_{21}$ in~(b). They were measured with a HP8722ES vector network analyser (VNA) in the frequency range from 50\,MHz to 40\,GHz at a power of -3\,dBm. The VNA was calibrated using the SOLT (short-open-load-trough) method with a calibration substrate (FormFactor CSR-8 100-250). Both devices were driven to the HRS prior to the frequency domain measurements and show a similar trend over the entire frequency range. Both are  insulating at low frequencies and become transparent at higher frequencies. This is due to the capacitance of the VCM stack, which acts as high pass filter. 

From the scatterings parameters, we determined the electrical charging time, by using the method described in \cite{Witzleben2021001_S}. The voltage at the device $V_\mathrm{DUT}(t)$ is the sum of the applied signal $V_\mathrm{P}(t)$ and the reflected signal $v_1^-(t)$, minus the transmitted signal $v_2^-(t)$:

\begin{equation}
V_\mathrm{DUT}(t) = V_\mathrm{P}(t) + v_1^-(t) - v_2^-(t).
\label{eq:vdut}
\end{equation}

\noindent An 250\,ps pulse with an amplitude of 0.8\,V was used as $V_\mathrm{P}(t)$, having a rise time of about 35\,ps (10\,\%-90\,\%). The reflected and transmitted signals were calculated by building the Fourier transform of $V_\mathrm{P}(t)$, multiplying it with $S_{11}$ for the reflection or with $S_{21}$ for the transmission, and finally building the inverse Fourier transform of this product:

\begin{equation}
v_1^-(t) = \mathcal{F}^{-1}\left(\mathcal{F}(v_1^+(t))\cdot S_{11}(f) \right),
\label{eq:dev_v-}
\end{equation}

\begin{equation}
v_2^-(t) = \mathcal{F}^{-1}\left(\mathcal{F}(v_1^+(t))\cdot S_{21}(f) \right).
\label{eq:dev_v2-}
\end{equation}

\begin{figure}[h]
\centering
\includegraphics[scale=1]{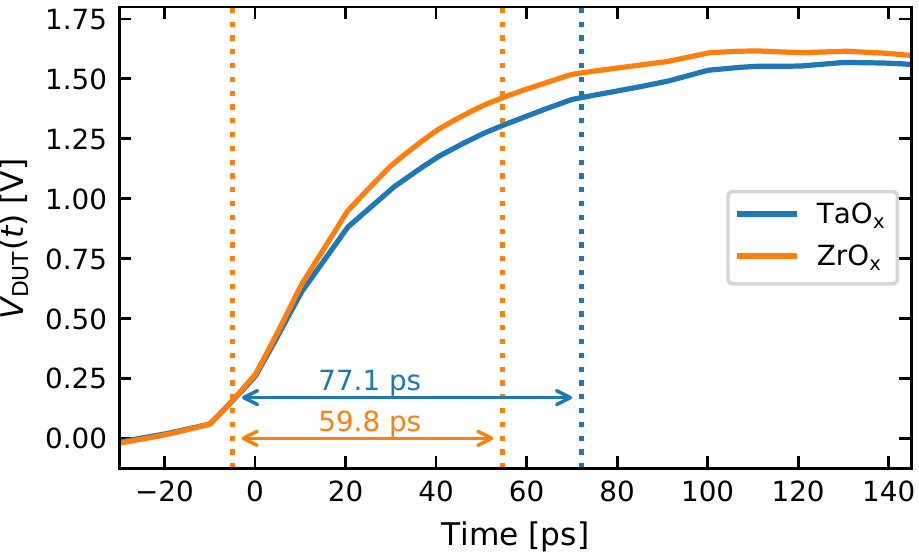}
\caption{Calculated voltage at the device $V_\mathrm{DUT}(t)$ after the application of 250\,ps pulse with an amplitude of 0.8\,V for the TaO$_\mathrm{x}$ (blue) and ZrO$_\mathrm{x}$-based (orange) device. The dotted vertical lines indicate the charging times (10\,\%-90\,\%).}
\label{sfig2}
\end{figure}

The results for $V_\mathrm{DUT}(t)$ of both devices are shown in Fig.~\ref{sfig2}, showing that both devices can be charged in less than 80\,ps (10\,\%-90\,\%). Please note that $V_\mathrm{DUT}(t)$ rises to twice the amplitude of $V_\mathrm{P}(t)$, which is due to the superposition of the incoming and reflected pulse. For this reason, we indicated always twice the pulse's amplitude in the main text.

\subsection{RESET failure of TaO$_\mathrm{x}$-based device at $V$ = 1.8\,V}

\begin{figure}[h]
\centering
\includegraphics[scale=1]{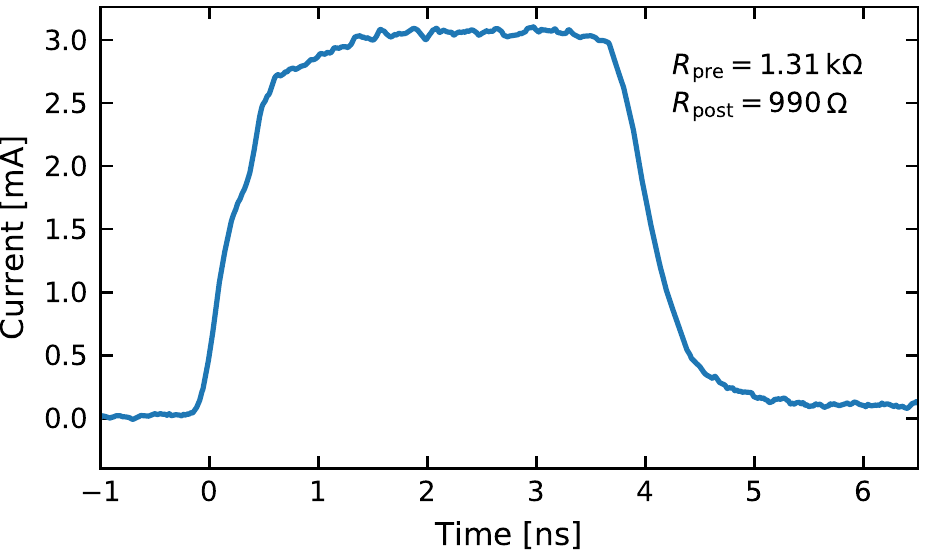}
\caption{Current response of the TaO$_\mathrm{x}$-based device to a 4\,ns with an amplitude of 1.8\,V.}
\label{sfig3}
\end{figure}

The current response of the TaO$_\mathrm{x}$-based device to a 4\,ns pulse with an amplitude of 1.8\,V is shown in Fig.~\ref{sfig3}. The devices was driven to a LRS of 1.31\,k$\Omega$ before the pulse's application. Although, a positive voltage pulse (applied to the active Pt~electrode) should drive the device to the HRS, the device's resistance decreased to 990\,$\Omega$, remaining in the LRS. Also, the current was significantly higher than during pulses with lower amplitude. At a pulse amplitude of 1.6\,V (see Fig.~\ref{fig2}(a) of the main text), the maximum current amounted to approx. 1.5\,mA and then decreased abruptly. In the case of 1.8\,V it remained constantly at approx.~3\,mA. This current increase can be explained by the occurrence of a unipolar SET event, which in this case prohibits a successful RESET operation. In a previous study on the RESET kinetics of a similar TaO$_\mathrm{x}$-based device, only a small linear dependence of the maximum current on the applied voltage was observed \cite{Marchewka2015002_S}. The observed current increase from 1.5\,mA to 3.0\,mA can, therefore, not be explained by the increase of voltage from 1.6\,V to 1.8\,V.

\subsection{RESET kinetics}

\begin{figure}[!p]
	\begin{subfigure}{\textwidth}
		\centering
		\includegraphics[scale = 1]{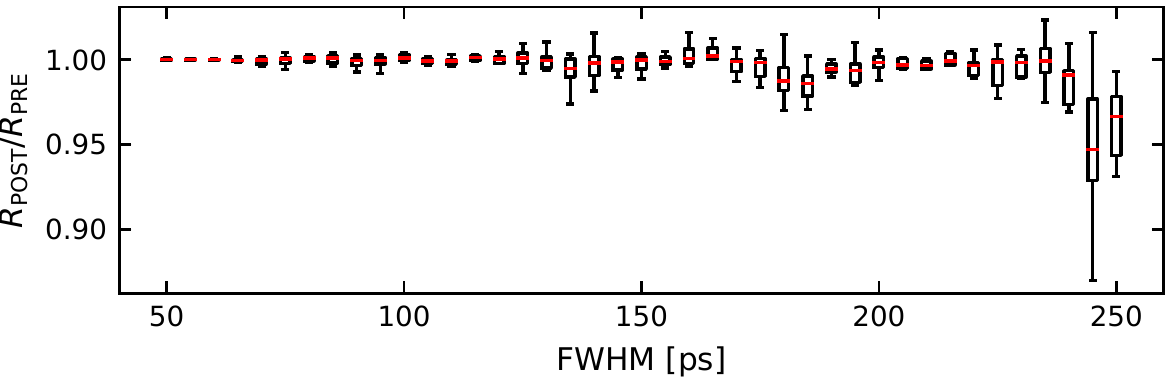}
		\subcaption{TaO$_\mathrm{x}$-based device, $V = 2.2$\,V.}
	\end{subfigure}\\
	\vspace{2mm}\\
	\begin{subfigure}{\textwidth}
		\centering
		\includegraphics[scale = 1]{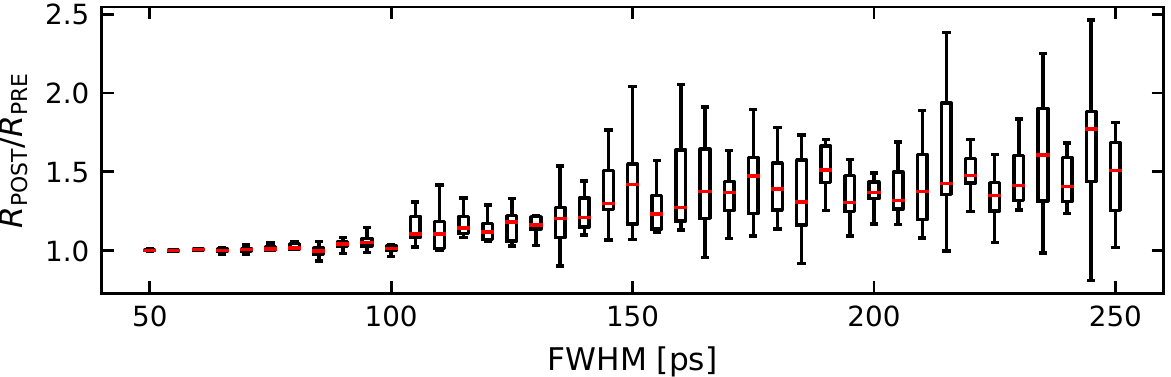}
		\subcaption{ZrO$_\mathrm{x}$-based device, $V = 1.6$\,V.}
	\end{subfigure}\\
	\vspace{2mm}\\
	\begin{subfigure}{\textwidth}
		\centering
		\includegraphics[scale = 1]{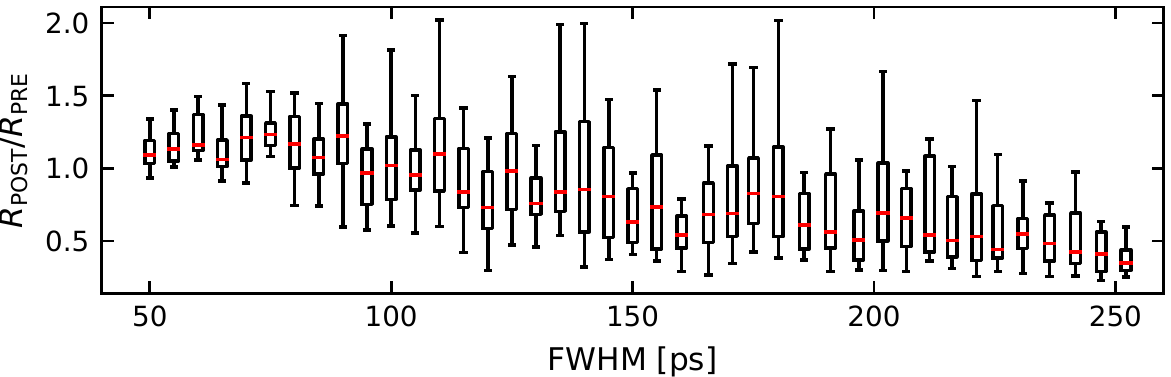}
		\subcaption{ZrO$_\mathrm{x}$-based device, $V = 2.2$\,V.}
	\end{subfigure}
	\caption{Boxplots of $R_\mathrm{POST}$/$R_\mathrm{PRE}$ values of the RESET kinetic measurements shown in Fig.~\ref{fig3}(c) of the main text. The red bar in the boxplot marks the median.}
	\label{fig:RESET_ratio}
\end{figure}

\begin{figure}[!p]
	\begin{subfigure}{\textwidth}
		\centering
		\includegraphics[scale = 1]{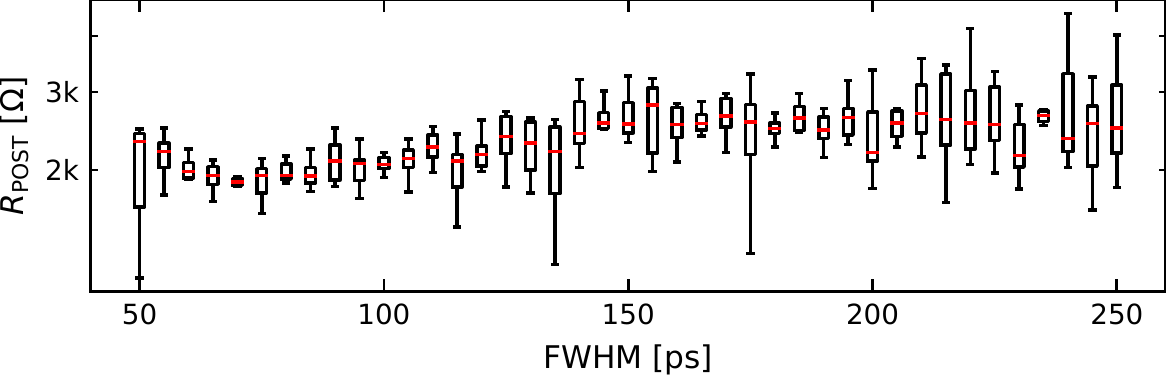}
		\subcaption{TaO$_\mathrm{x}$-based device, $V = 1.6$\,V.}
	\end{subfigure}\\
	\vspace{2mm}\\
	\begin{subfigure}{\textwidth}
		\centering
		\includegraphics[scale = 1]{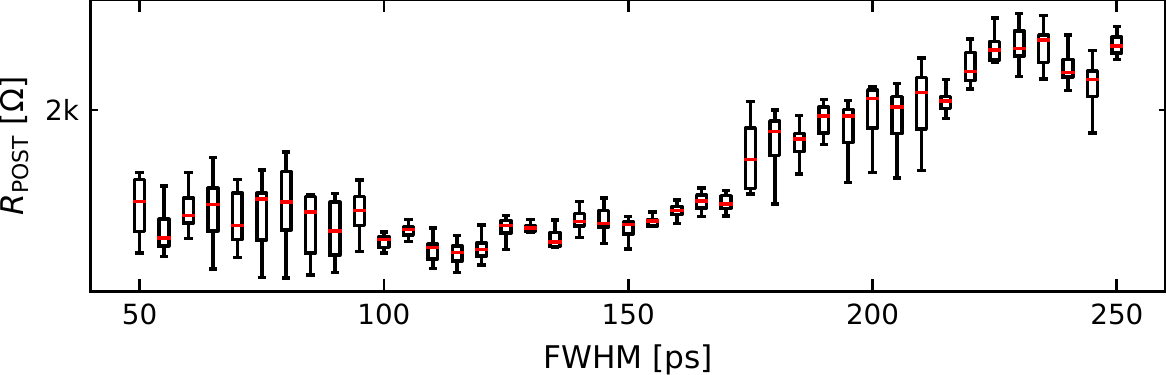}
		\subcaption{TaO$_\mathrm{x}$-based device, $V = 2.2$\,V.}
	\end{subfigure}\\
	\vspace{2mm}\\
	\begin{subfigure}{\textwidth}
		\centering
		\includegraphics[scale = 1]{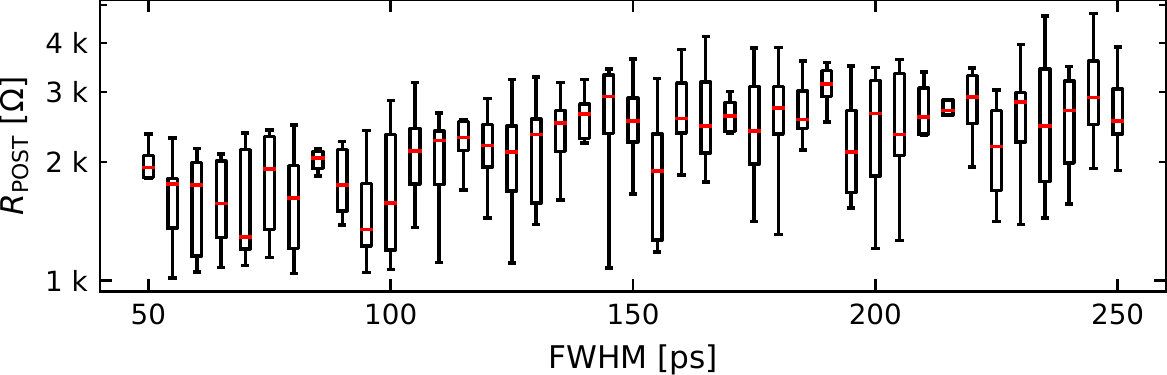}
		\subcaption{ZrO$_\mathrm{x}$-based device, $V = 1.6$\,V.}
	\end{subfigure}\\
	\vspace{2mm}\\
	\begin{subfigure}{\textwidth}
		\centering
		\includegraphics[scale = 1]{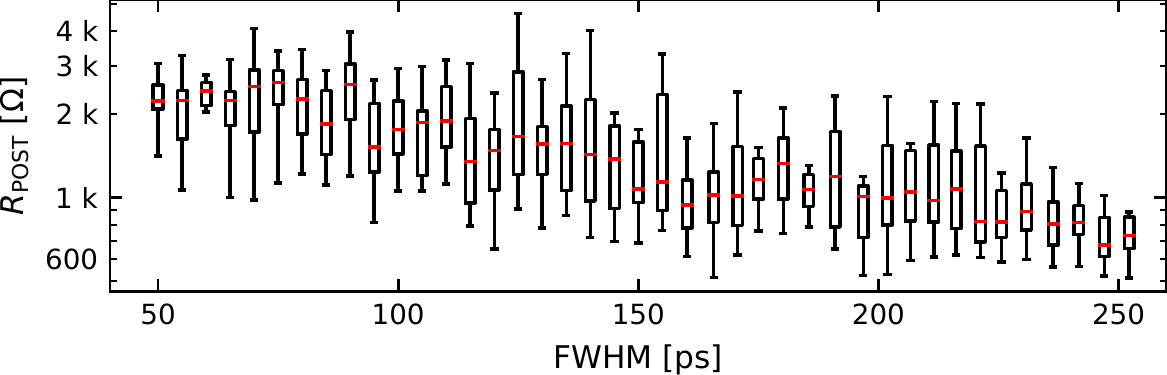}
		\subcaption{ZrO$_\mathrm{x}$-based device, $V = 2.2$\,V.}
	\end{subfigure}
	\caption{Boxplots of $R_\mathrm{POST}$ values of the RESET kinetic measurements shown in Fig.~\ref{fig3}(c) of the main text.}
	\label{fig:RESET_POST}
\end{figure}

Fig.~\ref{fig3}(c) of the main text shows the median values of the ratio $R_\mathrm{POST}$/$R_\mathrm{PRE}$ of the TaO$_\mathrm{x}$- and ZrO$_\mathrm{x}$-based device, which measured during the RESET kinetics in the time regime from 250\,ps to 50\,ps.  Two pulse amplitudes were used (1.6\,V and 2.2\,V). The boxplot representation was only shown for the TaO$_\mathrm{x}$-based device at 1.6\,V in Fig.~\ref{fig3}(b). The boxplot representation of the other three measurements is shown in Fig.~\ref{fig:RESET_ratio}. As the ratio $R_\mathrm{POST}$/$R_\mathrm{PRE}$ does not yield any information about the absolute resistance values, the boxplots of $R_\mathrm{POST}$ are shown in Fig.~\ref{fig:RESET_POST}. The $R_\mathrm{PRE}$ values were always in the range between 1.0\,k$\Omega$ and 2.5\,k$\Omega$.

\subsection{Unipolar SET -- Kinetics}

\begin{figure}[h]
	\begin{subfigure}{\textwidth}
		\centering
		\includegraphics[scale = 1]{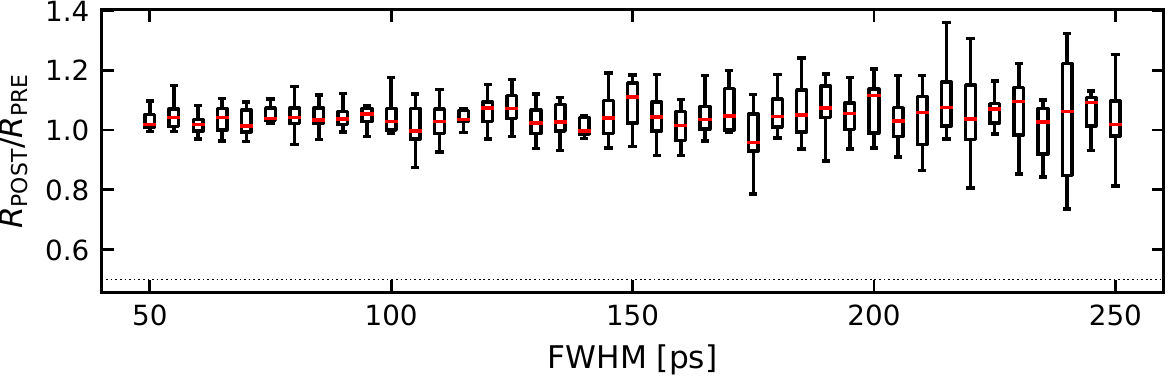}
		\subcaption{$V = 2.2$\,V.}
	\end{subfigure}\\
	\vspace{2mm}\\
	\begin{subfigure}{\textwidth}
		\centering
		\includegraphics[scale = 1]{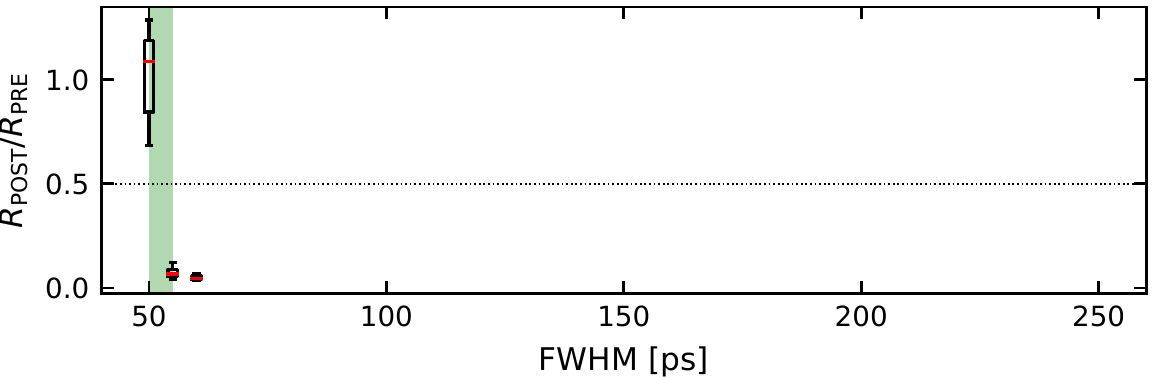}
		\subcaption{$V = 5.0$\,V.}
	\end{subfigure}
	\caption{Boxplots of $R_\mathrm{POST}$/$R_\mathrm{PRE}$ values of the unipolar SET kinetics measurements (shown in Fig.~\ref{fig4}(c) of the main text), conducted on the TaO$_\mathrm{x}$-based device.}
	\label{fig:SET_ratio_Ta}
\end{figure}

\begin{figure}[!h]
\begin{subfigure}{\textwidth}
		\centering
		\includegraphics[scale = 1]{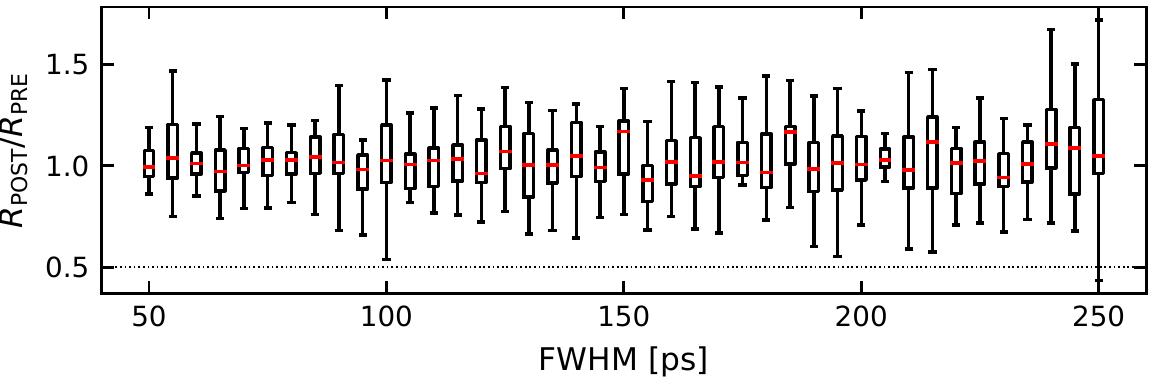}
		\subcaption{$V = 1.6$\,V.}
	\end{subfigure}\\
	\vspace{2mm}\\
	\begin{subfigure}{\textwidth}
		\centering
		\includegraphics[scale = 1]{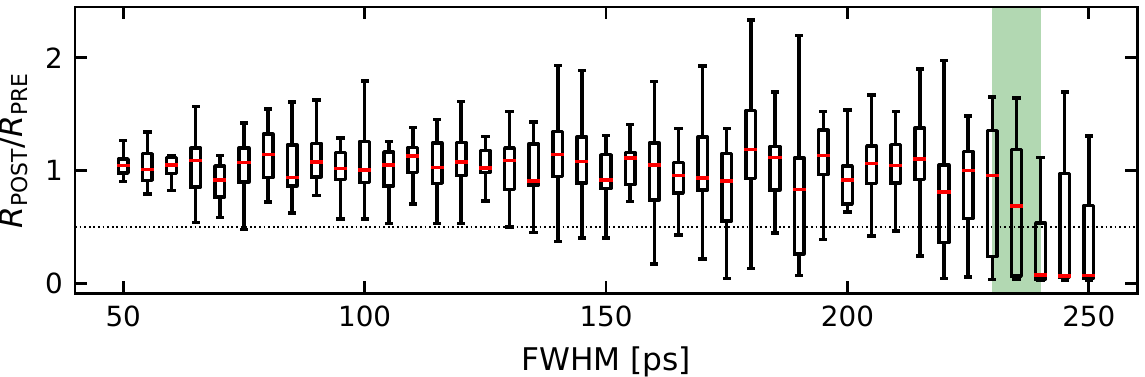}
		\subcaption{$V = 2.2$\,V.}
	\end{subfigure}\\
	\vspace{2mm}\\
	\begin{subfigure}{\textwidth}
		\centering
		\includegraphics[scale = 1]{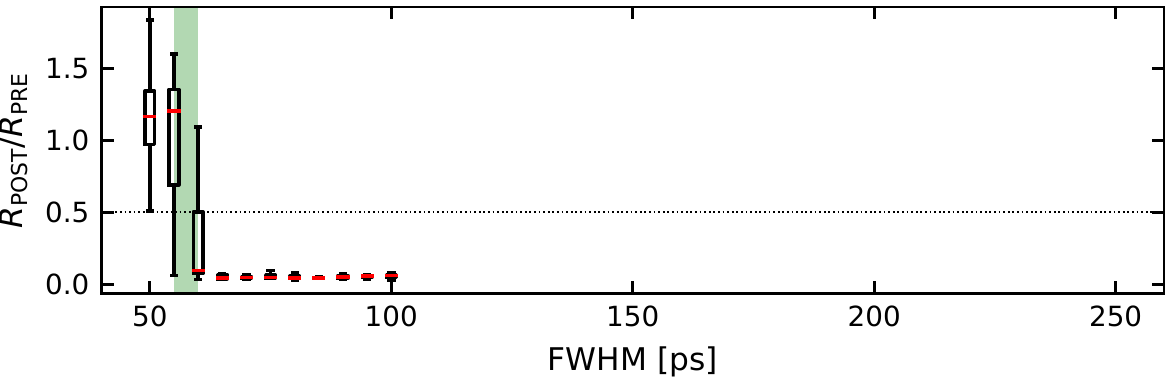}
		\subcaption{$V = 3.2$\,V.}
	\end{subfigure}\\
	\vspace{2mm}\\
	\begin{subfigure}{\textwidth}
		\centering
		\includegraphics[scale = 1]{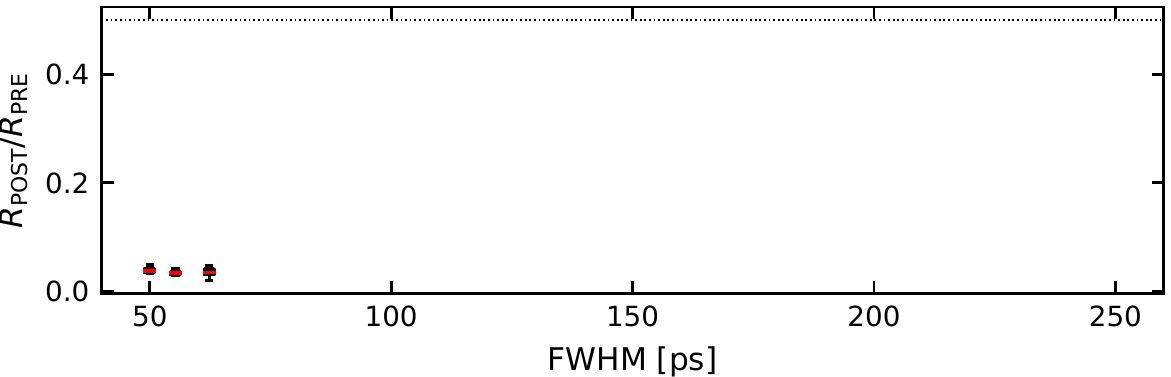}
		\subcaption{$V = 5.0$\,V.}
	\end{subfigure}
	\caption{Boxplots of $R_\mathrm{POST}$/$R_\mathrm{PRE}$ values of the unipolar SET kinetics measurements (shown in Fig.~\ref{fig4}(c) of the main text), conducted on the ZrO$_\mathrm{x}$-based device.}
	\label{fig:SET_ratio_Zr}
\end{figure}

\begin{figure}[h]
	\begin{subfigure}{\textwidth}
		\centering
		\includegraphics[scale = 1]{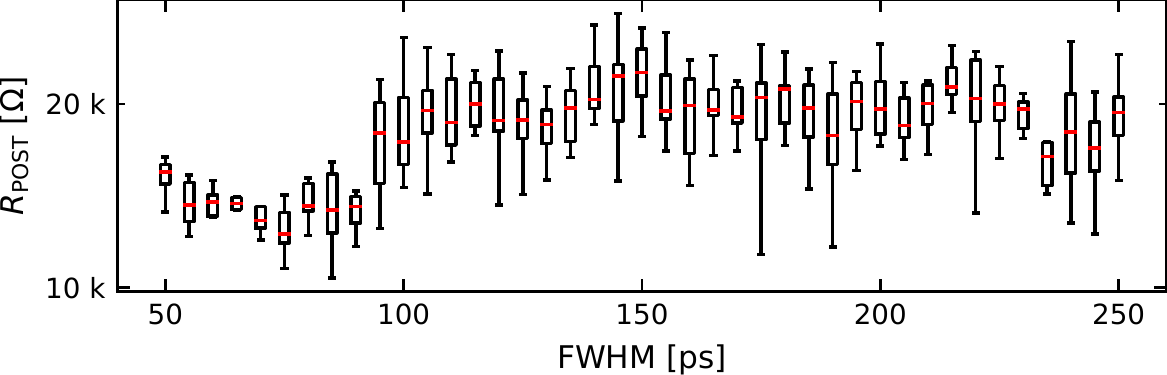}
		\subcaption{$V = 2.2$\,V.}
	\end{subfigure}\\
	\vspace{2mm}\\
	\begin{subfigure}{\textwidth}
		\centering
		\includegraphics[scale = 1]{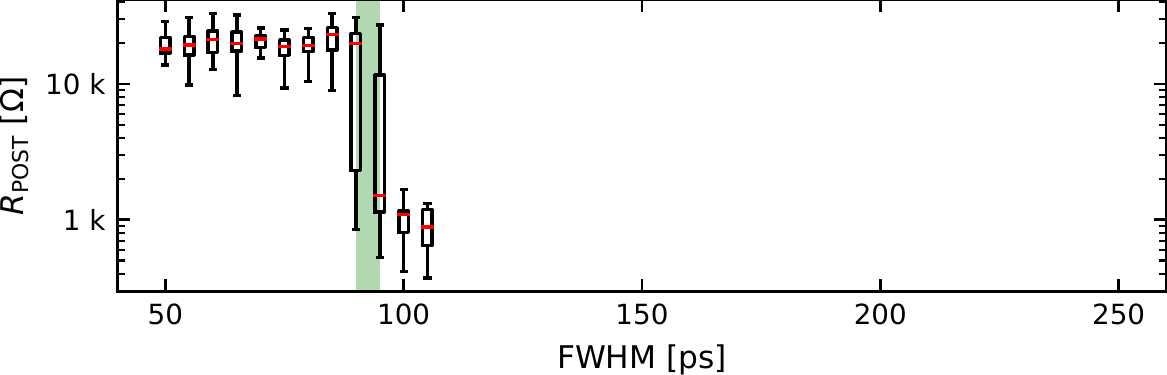}
		\subcaption{$V = 3.2$\,V.}
	\end{subfigure}\\
	\vspace{2mm}\\
	\begin{subfigure}{\textwidth}
		\centering
		\includegraphics[scale = 1]{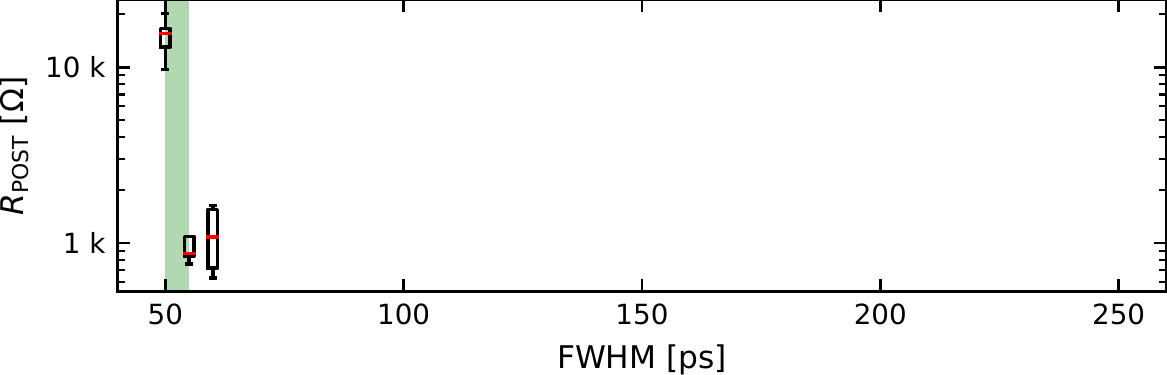}
		\subcaption{$V = 5.0$\,V.}
	\end{subfigure}
	\caption{Boxplots of $R_\mathrm{POST}$ values of the unipolar SET kinetics measurements (shown in Fig.~\ref{fig4}(c) of the main text), conducted on the TaO$_\mathrm{x}$-based device.}
	\label{fig:SET_ratio_Ta_POST}
\end{figure}

\begin{figure}[h]
\begin{subfigure}{\textwidth}
		\centering
		\includegraphics[scale = 1]{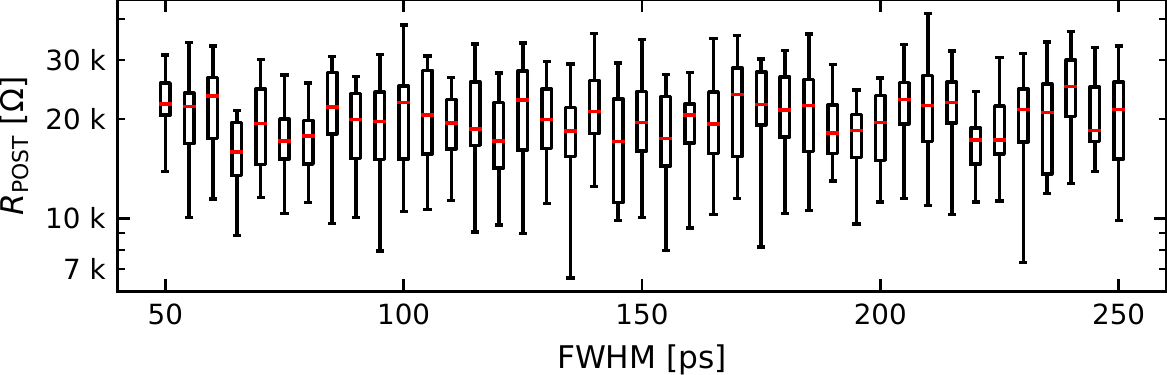}
		\subcaption{$V = 1.6$\,V.}
	\end{subfigure}\\
	\vspace{2mm}\\
	\begin{subfigure}{\textwidth}
		\centering
		\includegraphics[scale = 1]{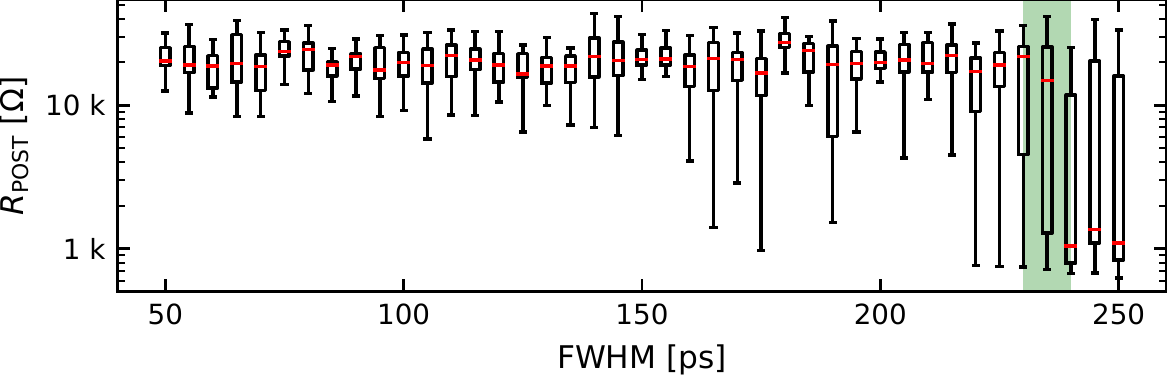}
		\subcaption{$V = 2.2$\,V.}
	\end{subfigure}\\
	\vspace{2mm}\\
	\begin{subfigure}{\textwidth}
		\centering
		\includegraphics[scale = 1]{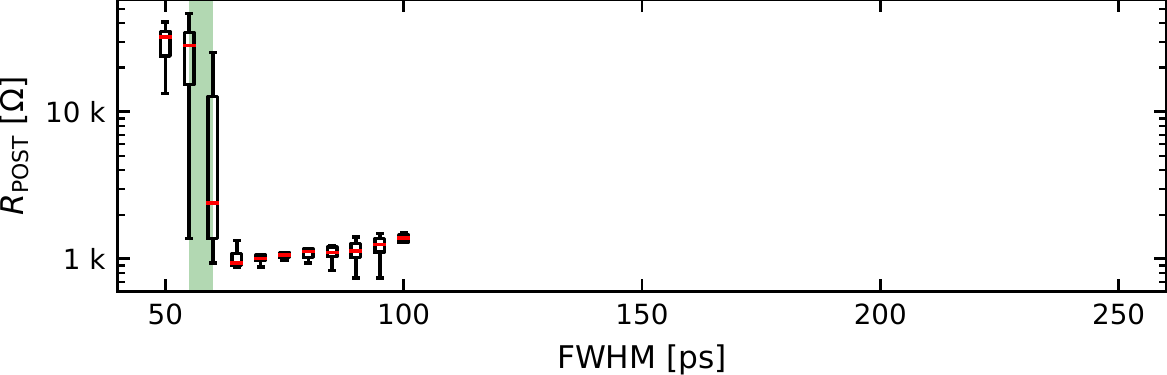}
		\subcaption{$V = 3.2$\,V.}
	\end{subfigure}\\
	\vspace{2mm}\\
	\begin{subfigure}{\textwidth}
		\centering
		\includegraphics[scale = 1]{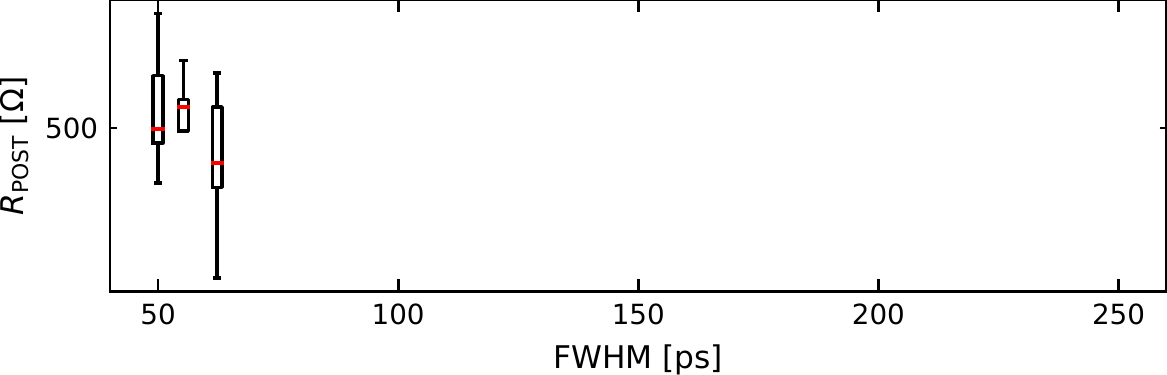}
		\subcaption{$V = 5.0$\,V.}
	\end{subfigure}
	\caption{Boxplots of $R_\mathrm{POST}$ values of the unipolar SET kinetics measurements (shown in Fig.~\ref{fig4}(c) of the main text), conducted on the ZrO$_\mathrm{x}$-based device.}
	\label{fig:SET_ratio_Zr_POST}
\end{figure}

\begin{figure}[!h]
	\begin{subfigure}{0.49\textwidth}
		\centering
		\includegraphics[scale = 1]{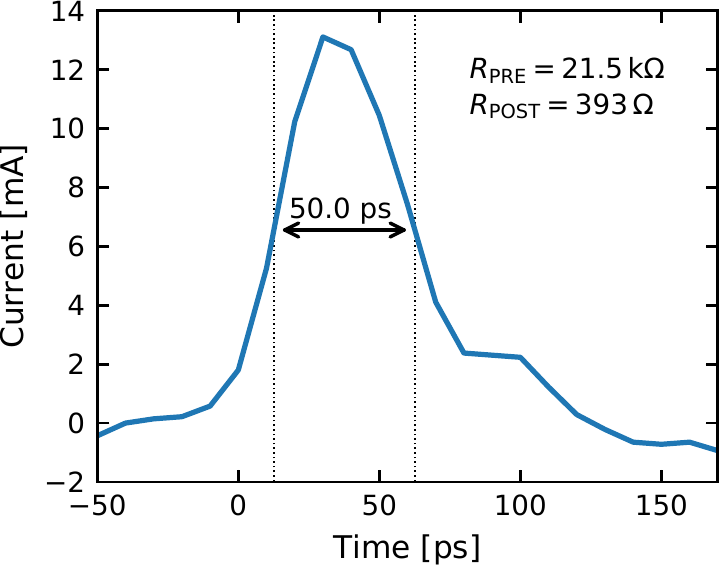}
		\subcaption{TaO$_\mathrm{x}$, $V = 7.0$\,V}
	\end{subfigure}
	\begin{subfigure}{0.49\textwidth}
		\centering
		\includegraphics[scale = 1]{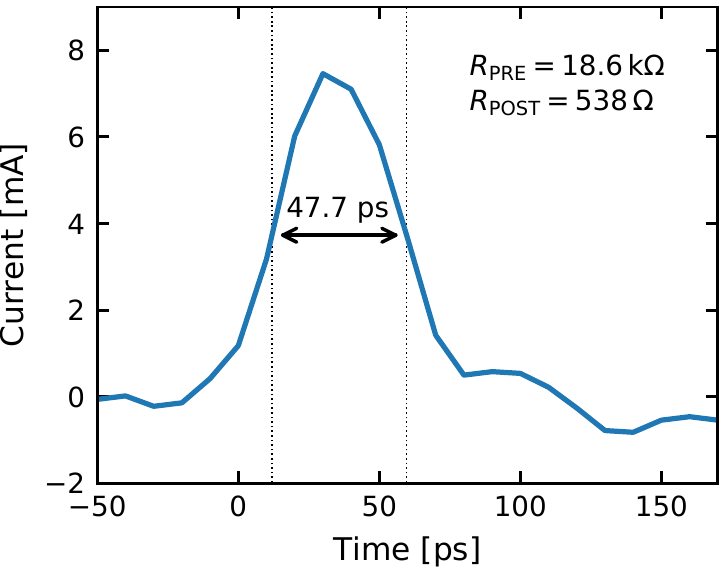}
		\subcaption{ZrO$_\mathrm{x}$, $V = 5.0$\,V}
	\end{subfigure}
	\caption{Current responses of the (a) TaO$_\mathrm{x}$- and (b) the ZrO$_\mathrm{x}$-based device to the application of a 50\,ps pulse. The devices' resistances before $R_\mathrm{PRE}$ and after $R_\mathrm{POST}$ the application of the pulse are indicated on the upper right, showing that the device switched from the HRS to the LRS during the pulse's application.}
	\label{fig:SET_50ps}
\end{figure}

Fig.~\ref{fig4}(c) of the main text shows the median values of the ratio $R_\mathrm{POST}$/$R_\mathrm{PRE}$ of the TaO$_\mathrm{x}$- and ZrO$_\mathrm{x}$-based device, which measured during the unipolar SET kinetics in the time regime from 250\,ps to 50\,ps. Four different pulse amplitudes were used, ranging from 1.6\,V to 5.0\,V. The boxplot representation was only shown for the TaO$_\mathrm{x}$-based device at 3.2\,V in Fig.~\ref{fig4}(b). The boxplot representation of the other measurements on the TaO$_\mathrm{x}$- and ZrO$_\mathrm{x}$-based device are shown in Fig.~\ref{fig:SET_ratio_Ta} and Fig.~\ref{fig:SET_ratio_Zr}, respectively. The boxplots of $R_\mathrm{POST}$ are shown in Fig.~\ref{fig:SET_ratio_Ta_POST} and Fig.~\ref{fig:SET_ratio_Zr_POST} for the TaO$_\mathrm{x}$- and the ZrO$_\mathrm{x}$-based device, respectively. The $R_\mathrm{PRE}$ values were always in the range between 10\,k$\Omega$ and 30\,k$\Omega$. The measurement cycle was repeated at least 10~times and until smooth lines for $R_\mathrm{POST}$/$R_\mathrm{PRE}$ were achieved. The number of conducted cycles are listed in Table~\ref{tab:no_cycles}. In Fig.~\ref{fig:SET_50ps} transient current responses to 50\,ps long pulses are shown for both devices. The devices switched from the HRS to the LRS during theses pulses.

\clearpage

\begin{table}[!h]
  \centering
  \caption{Number of conducted cycles}
    \begin{tabular}{c|c|c}
    Device & Voltage [V] & No. of cycles \\
    \midrule
    TaO$_\mathrm{x}$ & 2.2   & 100 \\
    TaO$_\mathrm{x}$ & 3.2   & 300 \\
    TaO$_\mathrm{x}$ & 5.0   & 100 \\
    ZrO$_\mathrm{x}$ & 1.6   & 300 \\
    ZrO$_\mathrm{x}$ & 2.2   & 300 \\
    ZrO$_\mathrm{x}$ & 3.2   & 500 \\
    ZrO$_\mathrm{x}$ & 5.0   & 100 \\
    \end{tabular}%
  \label{tab:no_cycles}%
\end{table}%

\subsection{Unipolar SET -- Endurance}

\begin{figure}[ht]
\begin{subfigure}{\textwidth}
		\centering
		\includegraphics[scale = 1]{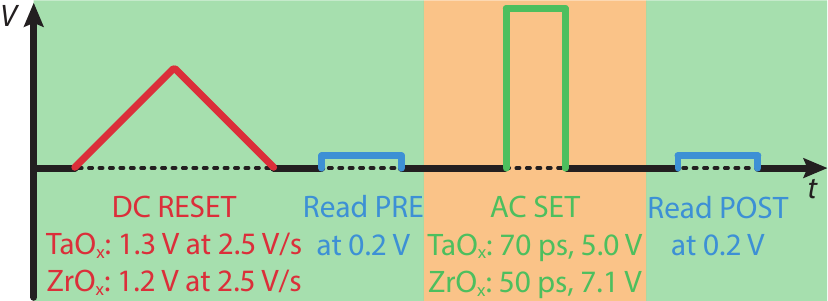}
		\subcaption{Sketch of endurance measurement.}
	\end{subfigure}	\\
	\vspace{2mm}\\
	\begin{subfigure}{0.49\textwidth}
		\centering
		\includegraphics[scale = 1]{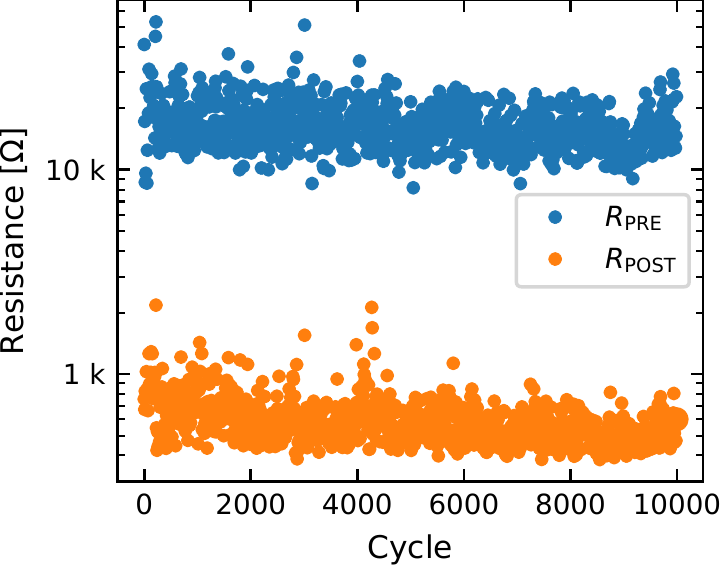}
		\subcaption{TaO$_\mathrm{x}$}
	\end{subfigure}
	\begin{subfigure}{0.49\textwidth}
		\centering
		\includegraphics[scale = 1]{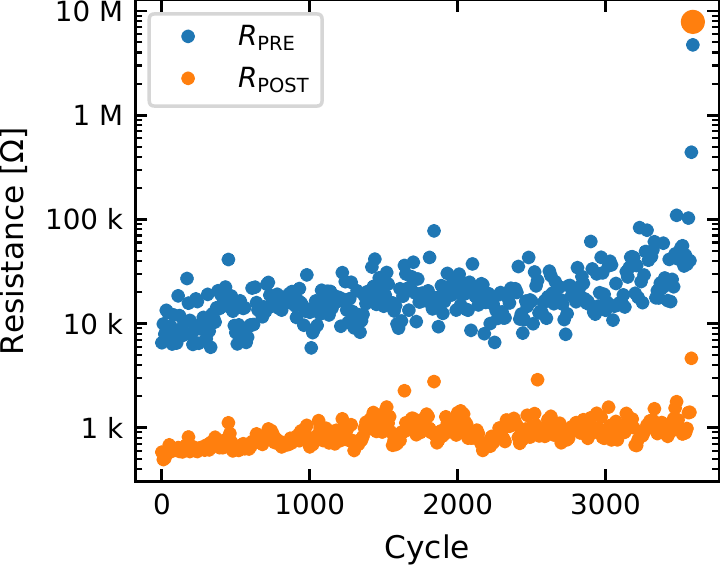}
		\subcaption{ZrO$_\mathrm{x}$}
	\end{subfigure}
	\caption{(a) Sketch of the pulse sequence for the endurance measurement. Resulting values for $R_\mathrm{PRE}$ (blue) and $R_\mathrm{POST}$ (orange) are shown in (b) and (c) for the TaO$_\mathrm{x}$- and ZrO$_\mathrm{x}$-based device, respectively.}
	\label{fig:SET_endurance}
\end{figure}

The measurement procedure to test the devices' endurance is sketched in Fig.~\ref{fig:SET_endurance}(a). All applied voltages were positive (including the read-out). At the beginning of each cycle, the device is driven into the HRS with a positive voltage sweep with an amplitude of 1.3\,V for the TaO$_\mathrm{x}$- and of 1.2\,V for the ZrO$_\mathrm{x}$-based device. The amplitude was reduced compared to the measurement in the main text to reduce the stress onto the devices. To reduce the measurement duration, the sweep rate was also accelerated from 0.5\,V/s to 2.5\,V/s. The devices were then driven into the LRS with a positive voltage pulse. For the TaO$_\mathrm{x}$-based device, the pulse width was chosen to 70\,ps and the pulse amplitude to 5.0\,V. For the ZrO$_\mathrm{x}$-based device, the pulse width was chosen to 50\,ps and the pulse amplitude to 7.1\,V. The devices' resistances were read at a voltage of 0.5\,V, before $R_\mathrm{PRE}$ and after $R_\mathrm{POST}$ the pulse's application.

The results are shown in Fig.~\ref{fig:SET_endurance}(b) and (c) for the TaO$_\mathrm{x}$- and ZrO$_\mathrm{x}$-based device, respectively. The measurement of the TaO$_\mathrm{x}$-based device was aborted after 10$^4$~cycles. The $R_\mathrm{PRE}$ and $R_\mathrm{POST}$ were always separated by an order of magnitude and an even higher endurance may be possible. The ZrO$_\mathrm{x}$-based device got stuck in the HRS after 3590~cycles. Until then, the $R_\mathrm{PRE}$ and $R_\mathrm{POST}$ were also separated by one order of magnitude. The device could, however, be driven back to the LRS with a negative voltage sweep and was again operational. Further adjustments of the pulse width and amplitude may result in an higher endurance, but this is beyond the scope of this study.

\singlespacing

\end{document}